\documentclass[pra,aps,twocolumn,amsfonts,showpacs,superscriptaddress]{revtex4-1}

\usepackage{epsfig}  
\usepackage{psfrag} 
\usepackage{amsmath}
\usepackage{amssymb} 
\usepackage{color} 
\usepackage{graphicx}
\usepackage{subfigure}

\newcommand{\be}{\begin{equation}} \newcommand{\ee}{\end{equation}}
\newcommand{\bea}{\begin{eqnarray}} \newcommand{\eea}{\end{eqnarray}}
 
\def\nn{\nonumber\\}

\begin{document}

\title{Discrete Symmetry Breaking Transitions Between Paired
  Superfluids}

\author{M. J. Bhaseen} \affiliation{Cavendish Laboratory, University
  of Cambridge, Cambridge, CB3 0HE, United Kingdom} \author{S. Ejima}
\affiliation{Institut f{\"u}r Physik,
  Ernst-Moritz-Arndt-Universit{\"a}t Greifswald, 17489 Greifswald,
  Germany} \author{F. H. L. Essler} \affiliation{The Rudolf Peierls
  Centre for Theoretical Physics, University of Oxford, Oxford, OX1
  3NP, United Kingdom} \author{H. Fehske} \affiliation{Institut
  f{\"u}r Physik, Ernst-Moritz-Arndt-Universit{\"a}t Greifswald, 17489
  Greifswald, Germany} \author{M. Hohenadler} \affiliation{Institut
  f\"ur Theoretische Physik und Astrophysik, Universit\"at W\"urzburg,
  97074 W\"urzburg, Germany} \author{B. D. Simons}
\affiliation{Cavendish Laboratory, University of Cambridge, Cambridge,
  CB3 0HE, United Kingdom}
\begin{abstract}
  We explore the zero-temperature phase diagram of bosons interacting
  via Feshbach resonant pairing interactions in one dimension. Using
  DMRG (Density Matrix Renormalization Group) and field theory
  techniques we characterize the phases and quantum phase transitions
  in this low-dimensional setting. We provide a broad range of
  evidence in support of an Ising quantum phase transition separating
  distinct paired superfluids, including results for the energy gaps,
  correlation functions and entanglement entropy. In particular, we
  show that the Ising correlation length, order parameter and critical
  properties are directly accessible from a ratio of the atomic and
  molecular two-point functions.  We further demonstrate that both the
  zero-momentum occupation numbers and the visibility are in
  accordance with the absence of a purely atomic superfluid
  phase.  We comment on the connection to recent studies of
  boson pairing in a generalized classical XY model.

\end{abstract}

\date{\today}


\maketitle

\section{Introduction}

The rapid progress in manipulating ultra cold atomic gases has led to
new approaches to strongly interacting quantum systems.  This includes
the properties of highly correlated states of matter, such as
Bose--Einstein condensates (BECs) \cite{Anderson:BEC,Ketterle:BEC},
Mott insulators \cite{Greiner:SI}, and supersolids
\cite{Baumann:Dicke}. It also allows access to the phase transitions
and crossovers between these fascinating phases.  In the last few
years, the BEC--BCS crossover
\cite{Eagles:Possible,Legget:Cooper,Nozieres:NSR,Leggett:QL} between a
molecular BEC and a Bardeen--Cooper--Schrieffer (BCS) pairing state,
has stimulated a wealth of experimental activity using fermionic atoms
\cite{Greiner:Emergence,Jochim:BECmol,Zwierlein:Observation,Regal:BCSBEC,Bourdel:BECBCS,Zwierlein:Condensation,Bartenstein:Collective,Ketterle:Making}. This
has been achieved through the use of Feshbach resonances, which enable
one to control the strength of pairing interactions using a magnetic
field. This has not only opened the door to central problems in
condensed matter physics, but also offers insights into the quantum
chemistry of molecule formation and chemical reactions
\cite{Ospelkaus:Reactions}.

In tandem with these advances, Feshbach resonances and molecule
formation have also been studied in bosonic systems
\cite{Inouye:Obs,Courteille:Obs,Donley:Atmol,Herbig:Prep,Durr:Diss,Xu:Form,Drummond:Coherent,Timmermans:Feshbach,Duine:Atmol}. Recent
experiments have been performed both in optical traps
\cite{Papp:Bragg,Pollack:Extreme,Navon:Dynamics} and in optical
lattices \cite{Thalhammer:Long}; for a review see
Ref.~\cite{Chin:Fesh}.  On the theoretical side, the BEC--BCS
``crossover'' problem for bosons has also been investigated in the
continuum limit \cite{Rad:Atmol,Romans:QPT,Radzi:Resonant} and on the
lattice
\cite{Dickerscheid:Feshbach,Sengupta:Feshbach,Rousseau:Fesh,Rousseau:Mixtures,Bhaseen:Feshising,Hohenadler:QPT}.
The problem differs markedly from the fermionic case since the
carriers themselves may Bose condense. This leads to the possibility
of an Ising quantum phase transition occurring between distinct paired
superfluids \cite{Rad:Atmol,Romans:QPT,Radzi:Resonant}.  The phases
are distinguished by the presence or absence of carrier condensation,
and the associated quantum phase transition involves discrete
${\mathbb Z}_2$ symmetry breaking. Closely related phases and quantum
phase transitions have also been observed in multicomponent fermion
systems
\cite{Wu:Competing,Lecheminant:Confinement,Capponi:Confinement,Roux:Spin},
and in the attractive Bose--Hubbard model with a restricted three
particle Hilbert space
\cite{Daley:Three,Daley:Threeerratum,Diehl:Observability,Diehl:QFTI,Diehl:QFTII,Bonnes:Pair,Bonnes:Unbinding}. There
are also magnetic analogues in quantum spin chains
\cite{Manmana:Spin1}.  More recently, the phenomenon of boson pairing
has also been explored in the context of a generalized classical XY
model with two competing harmonics in the periodic interactions
\cite{Shi:Pairing,Korshunov:Possible,Lee:Strings,Carpenter:genxy}. This
has led to the prediction of a novel phase diagram with unusual
criticality.

Motivated by the possibility of stabilizing pairing phases of bosons
\cite{Valatin:Collective,Coniglio:Condensation,Evans:Bosepairing,Nozieres:Paircond,Rice:Superbose,Kagan:Pairing}
in cold atomic gases, we recently investigated the bosonic Feshbach
resonance problem in one dimension (1D) \cite{Ejima:ID}. We employed
large scale DMRG (Density Matrix Renormalization Group)
\cite{White:DMRG,*White:Algorithms} and field theory techniques
\cite{Ejima:ID} in order to incorporate the effects of enhanced
quantum fluctuations in 1D.  Amongst our findings, we presented
compelling evidence for an Ising quantum phase transition separating
distinct superfluids \footnote{In this 1D setting we use the terms
  ``superfluid'' and ``condensate'' to indicate a phase with power-law
  correlations.}. The aim of the present manuscript is to shed further
light on this novel transition, and to provide a thorough discussion
of the superfluid phases in this 1D setting. In particular, we
describe a variety of methods to extract the Ising characteristics
from the gapless superfluid background.  We also provide a
quantitative finite-size scaling analysis of the zero-momentum
occupation numbers and the visibility. Our results are consistent with
the absence of a purely atomic superfluid phase with non-condensed
molecules, in contrast to the earlier suggestions of
Ref.~\cite{Rousseau:Fesh,Rousseau:Mixtures}.

The layout of this paper is as follows. In Sec.~\ref{Sect:Model} we
present the Hamiltonian under investigation and in Sec.~\ref{Sect:PD}
we discuss the phase diagram.  In Sec.~\ref{Sect:QFT} we describe the
associated field theory and gather our predictions for a variety of
local expectation values and correlation functions. We use these
results to characterize the different phases and to establish a
detailed comparison with DMRG. In Sec.~\ref{Sect:Mom} we provide a
quantitative account of the finite-size scaling of the zero-momentum
occupation numbers and the visibility. We contrast our results with
those of Refs.~\cite{Rousseau:Fesh,Rousseau:Mixtures}. In
Sec.~\ref{Sect:EE} we discuss the behavior of the entanglement entropy
and the emergence of Ising criticality at the transition between the
distinct paired superfluids.  We also discuss the behavior at the
superfluid--Mott insulator transitions. In Sec.~\ref{Sect:SR} we
describe the Ising scaling regime, and discuss a variety of ways to
extract the principal Ising characteristics.  This includes the Ising
order parameter and the correlation length using a finite-size scaling
analysis of the atomic and molecular correlation functions. We also
discuss the utility of a suitable ratio of the atomic and molecular
two-point functions for analyzing the Ising quantum phase transition.
We conclude in Sec.~\ref{Sect:Conc} and provide further directions for
research.

\section{Model}
\label{Sect:Model}

We consider the Hamiltonian
\begin{equation}
\begin{aligned}
H & =\sum_{i\alpha}\epsilon_\alpha n_{i\alpha}-\sum_i\sum_\alpha t_\alpha\left(b_{i\alpha}^\dagger b_{i+1 \alpha}^{\phantom \dagger}+{\rm H.c.} \right)\\
& \hspace{1cm} +\sum_{i\alpha\alpha^\prime} 
\frac{U_{\alpha\alpha^\prime}}{2}n_{i\alpha}(n_{i\alpha^\prime}-\delta_{\alpha\alpha^\prime})
+ \,H_{{\rm F}},
\label{atmolham}
\end{aligned}
\end{equation}
describing bosons, $b_{i\alpha}$, hopping on a 1D lattice with sites
$i$, where $\alpha=a,m$ labels atoms and molecules
\cite{Dickerscheid:Feshbach,Sengupta:Feshbach,Rousseau:Fesh,Rousseau:Mixtures,Bhaseen:Feshising,Hohenadler:QPT}. Here,
$\epsilon_\alpha$ are on-site potentials, $t_\alpha$ are nearest
neighbor hopping parameters, and $U_{\alpha\alpha^\prime}$ are
interactions.  We assume that molecule formation is described by the
s-wave Feshbach resonance term, 
\be 
H_{{\rm F}}= g\sum_i(m^\dagger_{i}
a_{i}a_{i}+{\rm H.c.}), 
\label{HF}
\ee where we denote $m_i\equiv b_{im}$ and $a_i\equiv b_{ia}$; for
recent work on the p-wave problem see
Refs.~\cite{Radzi:Spinor,Choi:Finite}.  This conversion implies that
the number of atoms and molecules are not separately conserved, but
the total, $N_{\rm T}\equiv \sum_{i}(n_{ia}+2n_{im})$, is
preserved. For simplicity, in writing Eq.~(\ref{atmolham}) we neglect
any effects of higher Bloch bands in optical lattices
\cite{Diener:comment,Dickerscheid:Reply,Buchler:Micro}.  In this
respect, the Hamiltonian (\ref{atmolham}) may be regarded as a lattice
regularization of the continuum models studied in
Refs.~\cite{Rad:Atmol,Romans:QPT,Radzi:Resonant}; see also
Refs.~\cite{Drummond:Coherent,Timmermans:Feshbach,Duine:Atmol}.  This
approach is very convenient for numerical simulations, and enables us
to investigate the superfluid transitions where lattice effects are
germane. It also allows us to make contact with previous quantum Monte
Carlo (QMC) simulations \cite{Rousseau:Fesh,Rousseau:Mixtures} and to
place the problem on a firmer footing.  As in the original works
\cite{Rad:Atmol,Romans:QPT,Radzi:Resonant}, we neglect the effects of
three body losses and finite molecular lifetimes.

In this manuscript we use DMRG on 1D systems with up to $L=512$ sites,
where we set the lattice spacing to unity and adopt energy units where
$t_a=1$.  We furthermore set $t_m=1/2$ throughout.  We work in the
canonical ensemble with the total density $\rho_{\rm T}=N_{\rm T}/L=2$
held fixed and allow up to five atoms and five molecules per site,
corresponding to a large Hilbert space of dimension $(6\times 6)^L$; 
for a discussion of the effects of changing the local Hilbert
  space dimension see Appendix~\ref{App:Trunc}.  With open (periodic)
boundary conditions we retain up to $m_\rho=2400$ ($m_\rho=3000$)
states in the density matrix in order to ensure that the discarded
weight is less than $1\times 10^{-10}$ ($1\times 10^{-8}$).

\section{Phase Diagram}
\label{Sect:PD}

As we discussed in Ref.~\cite{Ejima:ID}, the qualitative phase diagram
of the 1D lattice Hamiltonian (\ref{atmolham}) was previously
considered using QMC simulations
\cite{Rousseau:Fesh,Rousseau:Mixtures}.  In addition to delineating
the Mott insulating and superfluid phase boundaries, this work led to
intriguing predictions of superfluidity within the Mott phase, and an
additional superfluid phase not present in mean field theory
\cite{Rad:Atmol,Romans:QPT,Radzi:Resonant,Dickerscheid:Feshbach,Sengupta:Feshbach}. Although
we find very good quantitative agreement with many of the numerical
results \cite{Rousseau:Fesh,Rousseau:Mixtures}, these additional
predictions are at variance with our recent findings \cite{Ejima:ID}
which combine field theory with DMRG. This was also suggested by our
earlier studies using hardcore bosons
\cite{Bhaseen:Feshising,Hohenadler:QPT}.  It has recently been argued
that the absence of particle conservation hindered the interpretation
of these previous QMC simulations \cite{Eckholt:Comment}.  In this
manuscript we will further demonstrate that the use of momentum space
observables, including the zero-momentum occupation numbers and the
visibility, also complicated the interpretation of these earlier
finite-size QMC simulations.

\begin{figure}
\includegraphics[width=3.2in,clip=true]{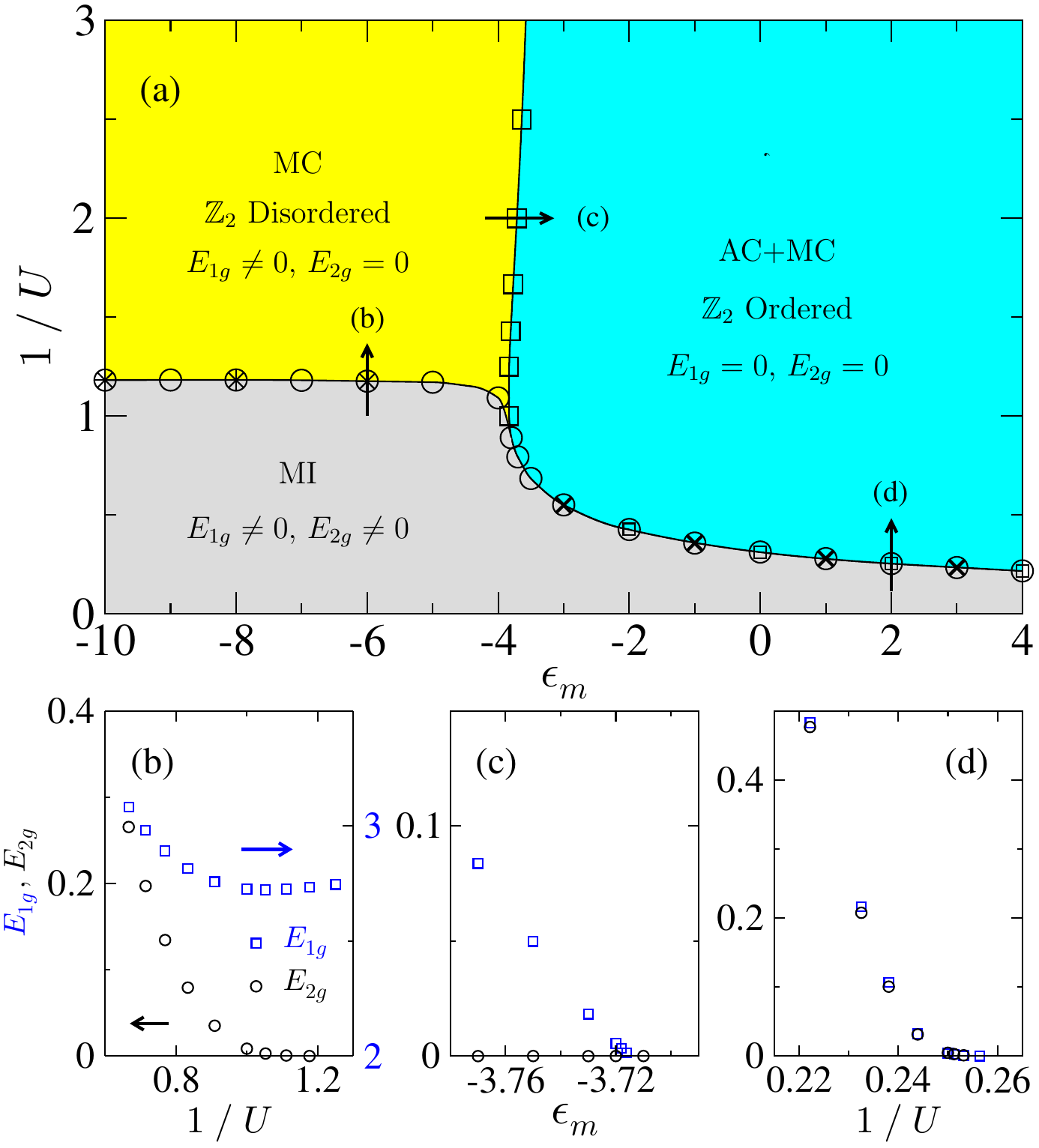}
\caption{(color online).  (a) Phase diagram of the 1D Hamiltonian
  (\ref{atmolham}) with total density $\rho_{\rm T}=N_{\rm T}/L=2$,
  showing a Mott insulator (MI), a molecular condensate (MC), and a
  coupled atomic plus molecular condensate (AC+MC). We use DMRG with
  up to $L=128$ sites and open boundary conditions. We choose
  parameters $\epsilon_a=0$, $U_{aa}/2=U_{mm}/2=U_{am}=g=U$, $t_a=1$,
  $t_m=1/2$, for comparison with Ref.~\cite{Rousseau:Fesh}. The
  squares and circles indicate the vanishing of the one-particle and
  two-particle gaps, $E_{1g}$ and $E_{2g}$, respectively, as
  $L\rightarrow\infty$.  The stars and crosses indicate where the
  molecular and atomic correlation exponents, $\nu_m$ and $\nu_a$
  reach $1/4$ in the MC and AC+MC phases respectively. These values
  correspond to a molecular KT transition and an atomic KT transition
  respectively.  The remaining panels (b), (c) and (d) show the
  variation of the extrapolated gaps, $E_{1g}$ and $E_{2g}$ with
  $L\rightarrow\infty$, on passing through the phase boundaries at the
  corresponding points in panel (a). In particular, we provide
  detailed evidence for an Ising quantum phase transition occurring
  between the MC and AC+MC phases.}
\label{Fig:PD}
\end{figure}

In order to put the problem
on a more stable platform, we present a section of the phase diagram
in Fig.~\ref{Fig:PD},
with parameters chosen for comparison with
Ref.~\cite{Rousseau:Fesh}. (Note that our conventions differ from
Ref.~\cite{Rousseau:Fesh} by a factor of $1/2$ in the interaction
terms so that double occupancy corresponds directly to
$U_{\alpha\alpha^\prime}$. Also, $\epsilon_m$ plays the role of their
detuning parameter, $D$, when $\epsilon_a=0$.)  The phase boundaries
shown in Fig.~\ref{Fig:PD} correspond to the vanishing of the
one-particle and two-particle excitation gaps, $E_{1g}\equiv
\mu_{1p}(L)-\mu_{1h}(L)$ and $E_{2g}\equiv \mu_{2p}(L)-\mu_{2h}(L)$
respectively, where the data are extrapolated to the thermodynamic
limit, $L\rightarrow\infty$. Here 
\begin{equation}
\begin{aligned}
\mu_{np}(L) & =\left[E_0(L,N_{\rm T}+n)-E_0(L,N_{\rm T})\right]/n,\\ 
\mu_{nh}(L) & =\left[E_0(L,N_{\rm T})-E_0(L,N_{\rm T}-n)\right]/n, 
\end{aligned}
\end{equation}
where $E_0(L,N)$ is the ground state energy for a system of size $L$
and a total number $N$ of atoms and molecules.  The phase diagram in
Fig.~\ref{Fig:PD} consists of three distinct phases: a Mott insulator
(MI) with gaps for both excitations, $E_{1g}\neq 0$ and $E_{2g}\neq
0$, a molecular condensate (MC) with a one-particle gap $E_{1g}\neq 0$
and $E_{2g}=0$, and a coupled atomic plus molecular condensate (AC+MC)
with $E_{1g}=0$ and $E_{2g}=0$. As we shall discuss more fully below,
the MC phase may be interpreted as a pairing phase of bosons in the
absence of atomic condensation. In contrast, the AC+MC phase has both
molecular and atomic condensation. In comparison to the qualitative
phase diagram presented in Ref.~\cite{Rousseau:Fesh}, inferred from
quantum Monte Carlo simulations on smaller system sizes, we find no
evidence for a single-component atomic superfluid phase co-existing
with non-condensed molecules. This is in accord with theoretical
expectations in higher dimensions, where atomic condensation is always
accompanied by molecular condensation
\cite{Rad:Atmol,Romans:QPT,Radzi:Resonant} provided the molecules are
present; in the extreme limit where $\epsilon_m\rightarrow\infty$,
occurring on the boundary of the AC+MC phase, the molecules are
explicitly excluded by the chemical potential as shown in
Fig.~\ref{Fig:Local}.  The conclusions of Ref.~\cite{Rousseau:Fesh}
have also come under scrutiny due to the additional claims of
superfluidity within the Mott phase
\cite{Bhaseen:Feshising,Eckholt:Comment}.  Here, however, our main
focus is on the character of the transition between the distinct MC
and AC+MC superfluids. In the subsequent discussion we will begin with
symmetry arguments and field theory predictions before turning to a
comparison with DMRG.

\section{Field Theory Description}
\label{Sect:QFT}
A heuristic way to understand the possibility of an Ising quantum
phase transition between the distinct MC and AC+MC superfluids is via
the generic number-phase relationships,
$a\sim\sqrt{\rho_a}\,e^{i\vartheta_a}$ and
$m\sim\sqrt{\rho_m}\,e^{i\vartheta_m}$, where $\rho_a$ and $\rho_m$
are the average atomic and molecular densities
respectively. Substituting these expressions into (\ref{atmolham}),
the Feshbach term (\ref{HF}) takes the form \cite{Radzi:Resonant} \be
H_{\rm F}\sim 2g\rho_a\sqrt{\rho_m}\,\cos(\vartheta_m-2\vartheta_a).
\ee Minimizing this interaction locks the phases of the atomic and
molecular condensates according to the relationship \be
\vartheta_m-2\vartheta_a=\pm \pi, \ee where for simplicity we assume
$g>0$. We see that the phases are locked, but only modulo $\pi$, and
this gives rise to the possibility of a discrete symmetry breaking
${\mathbb Z}_2$ transition between Feshbach coupled superfluids.
Denoting $\vartheta_m\equiv\vartheta$, one may recast the number-phase
relationships in the form \cite{Radzi:Resonant}
\begin{equation}
m\sim
\sqrt{\rho_m}\,e^{i\vartheta}, \quad a\sim\phi\,e^{i\vartheta/2},
\label{amising}
\end{equation}
where the Feshbach locking is explicitly enforced and $\phi\sim
\sqrt{\rho_a}e^{\mp i\pi/2}$ plays the role of an Ising degree of
freedom. The decomposition (\ref{amising}) will play a central role in
the subsequent analysis and allows one to gain a handle on the
correlation functions and the principal features of the phase diagram.

An alternative way to understand the possibility of an Ising quantum
phase transition between the MC and AC+MC phases is via the symmetry
of the Hamiltonian (\ref{atmolham}) under ${\rm U}(1)\times {\mathbb
  Z}_2$ transformations:
\begin{equation}
  m\rightarrow e^{i\theta}m, \quad a\rightarrow e^{i(\theta/2\pm
    \pi)}a, 
\end{equation}
where $\theta\in {\mathbb R}$. Before discussing the problem in 1D,
where continuous symmetry breaking is absent, we first consider the
behavior in higher dimensions
\cite{Rad:Atmol,Romans:QPT,Radzi:Resonant}.  In this case, the
molecular condensate (MC) phase has $\langle m\rangle\neq 0$ and
$\langle a\rangle=0$. This only breaks the ${\rm U}(1)$ symmetry, and
leaves the ${\mathbb Z}_2$ symmetry, $a\rightarrow -a$, unbroken. This
corresponds to an Ising degree of freedom in the disordered phase,
coexisting with molecular superfluidity. In contrast, the coupled
atomic plus molecular condensate (AC+MC) phase has $\langle m\rangle
\neq 0$ {\em and} $\langle a\rangle\neq 0$. This breaks the ${\rm
  U}(1)\times {\mathbb Z}_2$ symmetry completely and corresponds to a
${\mathbb Z}_2$ ordered Ising degree of freedom, coexisting with
atomic and molecular superfluidity. Returning to the present 1D
problem, where continuous ${\rm U}(1)$ symmetry breaking is
prohibited, the formation of expectation values for $\langle a\rangle$
and $\langle m\rangle$ is excluded. Instead, superfluidity is
characterized by power-law correlations, and the nature of the phases
and quantum phase transitions in Fig.~\ref{Fig:PD} requires further
examination.

Due to the ${\rm U}(1)\times {\mathbb Z}_2$ symmetry of the
Hamiltonian (\ref{atmolham}), the low energy Lagrangian of the MC to
AC+MC transition is given by ${\mathcal L}={\mathcal
  L}_\vartheta+{\mathcal L}_{\phi}+{\mathcal L}_{\vartheta\phi}$
\cite{Lee:Bosefeshbach,Radzi:Resonant}, where
\begin{equation} {\mathcal L}_\vartheta=
  \frac{K_\vartheta}{2}\left[c_\vartheta^{-2}(\partial_\tau\vartheta)^2
    +(\partial_x\vartheta)^2\right]
\label{scalar}
\end{equation}
is a free bosonic field and 
\begin{equation} {\mathcal
    L}_\phi=\frac{K_\phi}{2}\left[c_\phi^{-2}(\partial_\tau\phi)^2
    +(\partial_x\phi)^2\right]-{\mathcal M}\phi^2+\lambda\phi^4
\label{lphi}
\end{equation}
is an Ising model in the soft-spin $\phi^4$ representation.  The
coupling between the two sectors, ${\mathcal
  L}_{\vartheta\phi}=i\phi^2\partial_\tau\vartheta/2$, has a form
similar to a Berry phase \cite{Lee:Bosefeshbach,Radzi:Resonant}. A
closely related action also arises for tunnelling between quantum
wires \cite{Sitte:Emergent}.  In the following we will neglect the
contribution ${\mathcal L}_{\vartheta\phi}$ and explore the
consequences of the reduced action.  Sufficiently far away from the
transition this can be justified by a mean-field decoupling,
${\mathcal L}_{\vartheta\phi}\sim
i\langle\phi\rangle^2\partial_\tau\vartheta/2$, which reduces the
additional interaction to a total derivative term, which can be
neglected. The simplified action is therefore expected to provide a
good description of the proximate phases.  Near the quantum phase
transitions, this cannot be neglected {\em a priori}, and ${\mathcal
  L}_{\vartheta\phi}$ may change the behavior on very large length
scales and in other regions of the phase diagram
\cite{Sitte:Emergent}.  However, all of our findings are consistent
with expectations based on ${\mathcal L}_\vartheta+{\mathcal L}_\phi$
only.  The parameters $K_\vartheta$, $c_\vartheta$, $K_\phi$,
$c_\phi$, $\eta$, $\lambda$, are related to the coefficients of the
Hamiltonian (\ref{atmolham}), but the details need not concern us
here.  In this field theory approach, the atoms and molecules are
described by the semiclassical number-phase relations given in
Eq.~(\ref{amising}).  In the subsequent discussion we will explore the
ramifications of this correspondence in 1D, both for local observables
and correlation functions.  For complementary work using the Bethe
Ansatz and bosonization see also
Refs.~\cite{Gurarie:1DBFBR,*Gurarie:1DBF,Lee:1DBF}.

\subsection{Local Expectation Values}
\begin{figure}
\includegraphics[width=3.2in,clip=true]{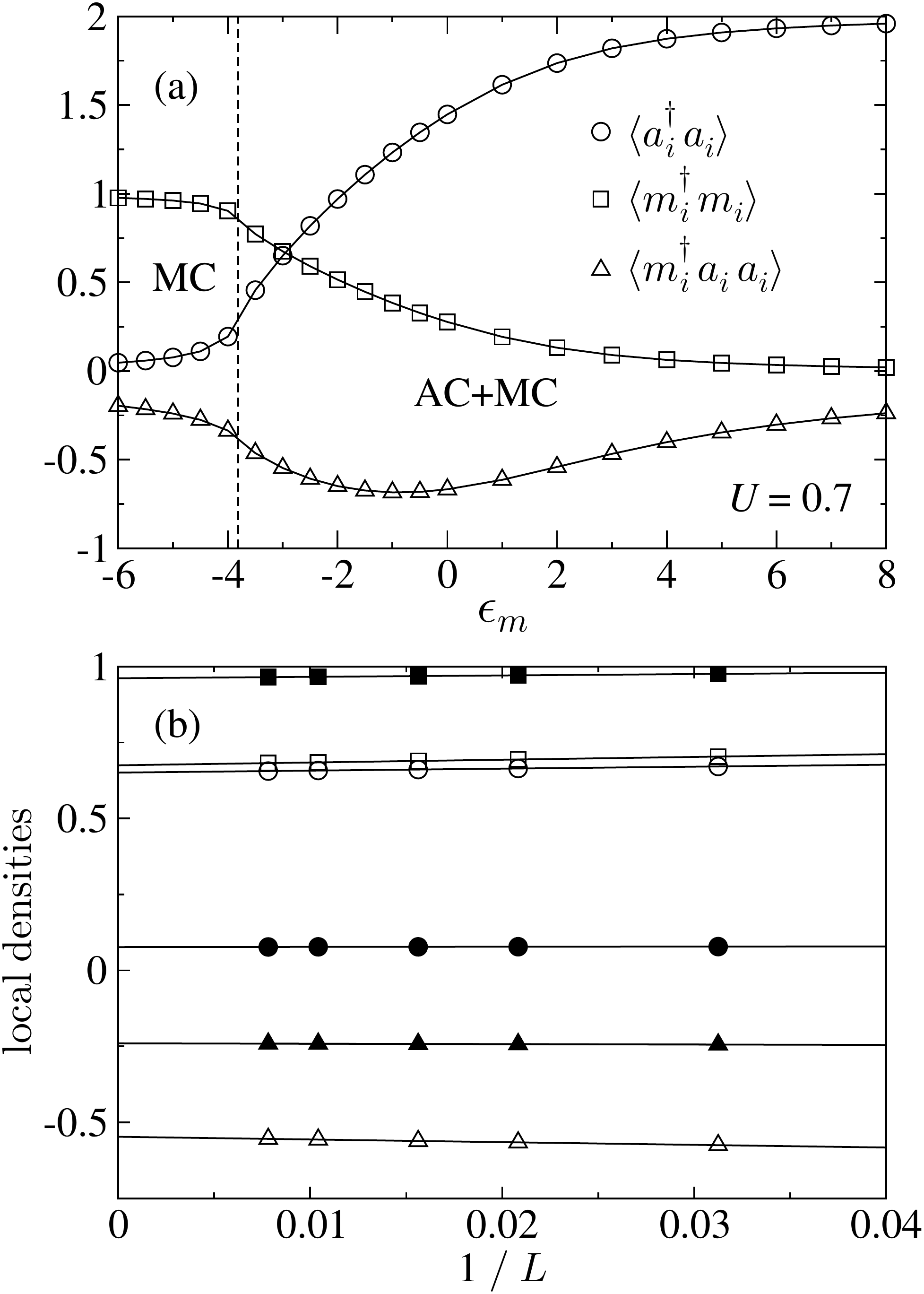}
\caption{(a) Local expectation values obtained by DMRG for the 1D
  Hamiltonian (\ref{atmolham}) with open boundaries and up to $L=128$
  sites.  We use the same parameters as in Fig.~\ref{Fig:PD} and set
  $U=0.7$. We plot the local density of atoms $\langle
  a^\dagger_ia_i\rangle$ (circles), molecules $\langle
  m^\dagger_im_i\rangle$ (squares), and the expectation value of the
  Feshbach conversion term $\langle m^\dagger_ia_ia_i\rangle$
  (triangles), evaluated at the system mid-point, $i=L/2$. All of
  these quantities are generically non-zero on both sides of the MC to
  AC+MC transition as indicated by the dashed line. In the
  limit of large positive (negative) detuning $\epsilon_m$ we have
  mainly atoms (molecules) with a density determined by the canonical
  ensemble constraint $\rho_{\rm T}=2$.  (b) Corresponding finite-size
  data and linear extrapolation as a function of $1/L$ for
  $\epsilon_m=-3$ (open) and $\epsilon_m=-5$ (filled).}
\label{Fig:Local}
\end{figure}

An immediate consequence of the decomposition (\ref{amising}) is that
the densities of atoms and molecules \be \langle
m^\dagger(x)m(x)\rangle \sim \rho_m,\quad \langle
a^\dagger(x)a(x)\rangle \sim \langle\phi^2\rangle, \ee are generically
non-zero in both the MC and AC+MC phases. This is supported by our
DMRG results as shown in Fig.~\ref{Fig:Local}(a). These are
extrapolated from the finite-size data to $L\rightarrow\infty$, as
indicated in Fig.~\ref{Fig:Local}(b).
In the limit of large positive detuning with
$\epsilon_m\rightarrow\infty$ we have mainly atoms as one would
naively expect and $\langle a^\dagger(x)a(x)\rangle\sim 2$. Likewise,
in the limit of large negative detuning, corresponding to
$\epsilon_m\rightarrow -\infty$, we have mainly molecules and $\langle
m^\dagger(x)m(x)\rangle\sim 1$. These limiting densities are
consistent with working in the canonical ensemble with $\rho_{\rm
  T}=\sum_i(n_{ia}+2n_{im})/L=2$ held fixed.

In addition to these local densities the local expectation value of
the Feshbach conversion term, \be \langle m^\dagger(x)a(x)a(x)\rangle
\sim \sqrt{\rho_m}\, \langle \phi^2\rangle\neq 0,
\label{feshev}
\ee is non-zero. It exhibits true long range order, even in this
low-dimensional setting. This is a consequence of the relevance of the
Feshbach term in the renormalization group sense. Our numerical
results in Fig.~\ref{Fig:Local} show that this quantity is indeed
finite. In particular, this confirms the locking of the phases of the
atomic and molecular condensates (modulo $\pi$) on {\em both} sides of
the transition. However, due to the symmetry under $a\rightarrow -a$,
the expectation value (\ref{feshev}) is naively insensitive to the
Ising transition itself, as may be seen in
Fig.~\ref{Fig:Local}. Further insight into this quantum phase
transition and the proximate phases is more readily obtained from
correlation functions. We will explore this in more detail below.

\subsection{Green's Functions and Pairing Correlations}
\label{Sec:GF}

The nature of the MC and AC+MC phases shows up most clearly in the
atomic and molecular Green's functions, $\langle
a^\dagger(x)a(0)\rangle$ and $\langle m^\dagger(x)m(0)\rangle$, and
the pairing correlations $\langle
a^\dagger(x)a^\dagger(x)a(0)a(0)\rangle$.  Their spatial dependence is
dictated by the correlations of the underlying Ising model in
Eq.~(\ref{lphi}), and we address each phase in turn.

\subsubsection{${\mathbb Z}_2$ Disordered MC Phase}

As follows from the decomposition (\ref{amising}), the molecular
Green's function \be \langle m^\dagger(x)m(0)\rangle\propto \langle
e^{-i\vartheta(x)}\
e^{i\vartheta(0)}\rangle\sim\left(\frac{a_0}{x}\right)^{\nu_m},
\label{mmdis}
\ee decays as a power-law, where the correlation exponent,
$\nu_m=1/(2\pi K_\vartheta)$ varies throughout the phase
diagram, and $a_0$ is a short-distance cutoff. 
In contrast, in the ${\mathbb Z}_2$ disordered MC phase, the
atomic Green's function \be\langle a^\dagger(x)a(0)\rangle \propto
\langle \phi(x) e^{-i\frac{\vartheta(x)}{2}}
\phi(0)e^{i\frac{\vartheta(0)}{2}}\rangle \nn
\sim\left(\frac{a_0}{x}\right)^\frac{\nu_m}{4}{\rm K}_0(x/\xi),
\label{aadis}
\ee decays exponentially, where $\xi$ is the Ising correlation length.
Here we use the hard-spin fermionic representation of the Ising model
to write \cite{Wu:Spin} \be \langle \phi(x) \phi(0)\rangle \sim\ {\rm
  K}_0(x/\xi), \ee where ${\rm K}_0$ is a modified Bessel function.
On the other hand, {\em pairs} of atoms exhibit power-law correlations
\be \langle a^\dagger(x)a^\dagger(x)a(0)a(0)\rangle \sim
\left(\frac{a_0}{x}\right)^{\nu_b}, \ee where the exponent
$\nu_b=\nu_m$ for these atomic bilinears coincides with the molecular
exponent in Eq.~(\ref{mmdis}).  That is to say, the MC phase is a
pairing phase of bosons without power-law atomic condensation
\cite{Valatin:Collective,Coniglio:Condensation,Evans:Bosepairing,Nozieres:Paircond,Rice:Superbose,Kagan:Pairing}.

\begin{figure}
\includegraphics[width=3.2in,clip=true]{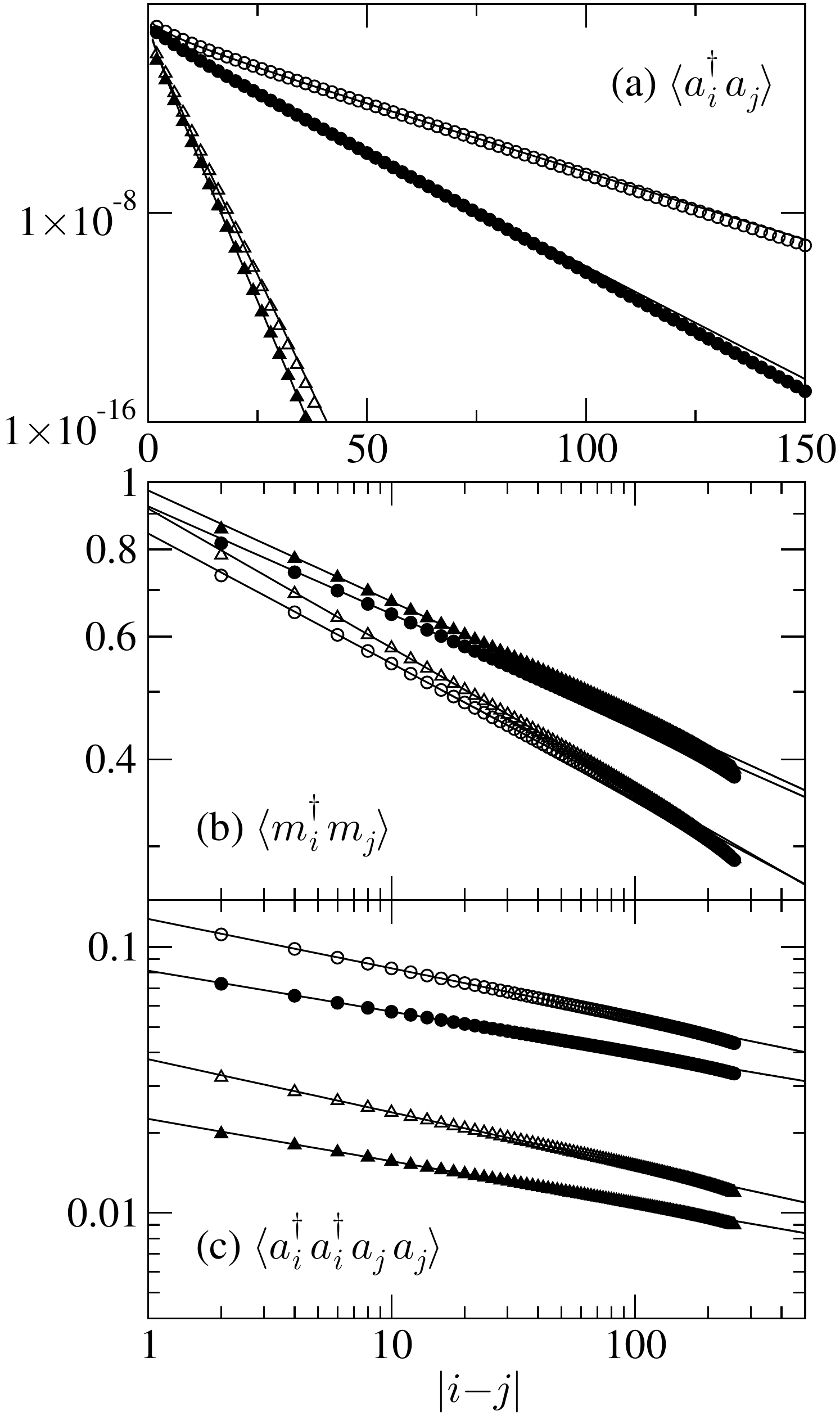}
\caption{Correlation functions in the ${\mathbb Z}_2$ disordered MC
  phase obtained by DMRG on the 1D Hamiltonian (\ref{atmolham}) with
  $L=512$ and open boundaries. Here and throughout the manuscript we
  consider sites displaced around the system mid-point in order to
  minimize boundary effects. We use the same parameters as in
  Fig.~\ref{Fig:PD} and set $U=0.7$ (open), $U=0.5$ (filled),
  $\epsilon_m=-4$ (circles) and $\epsilon_m=-6$ (triangles).  (a)
  Atomic Green's function $\langle a^\dagger_ia_j\rangle$ showing
  exponential decay. (b) Molecular Green's function $\langle
  m^\dagger_im_j\rangle$ showing power-law behavior.  (c) Bilinears of
  atoms $\langle a^\dagger_ia^\dagger_ia_ja_j\rangle$ showing
  power-law behavior with the same exponent as the molecular Green's
  function in panel (b); see Fig.~\ref{Fig:nummc}.  This establishes
  the MC phase as a pairing phase of atoms without power-law atomic
  condensation.}
\label{Fig:discorr}
\end{figure}

\begin{figure}
\includegraphics[width=3.2in,clip=true]{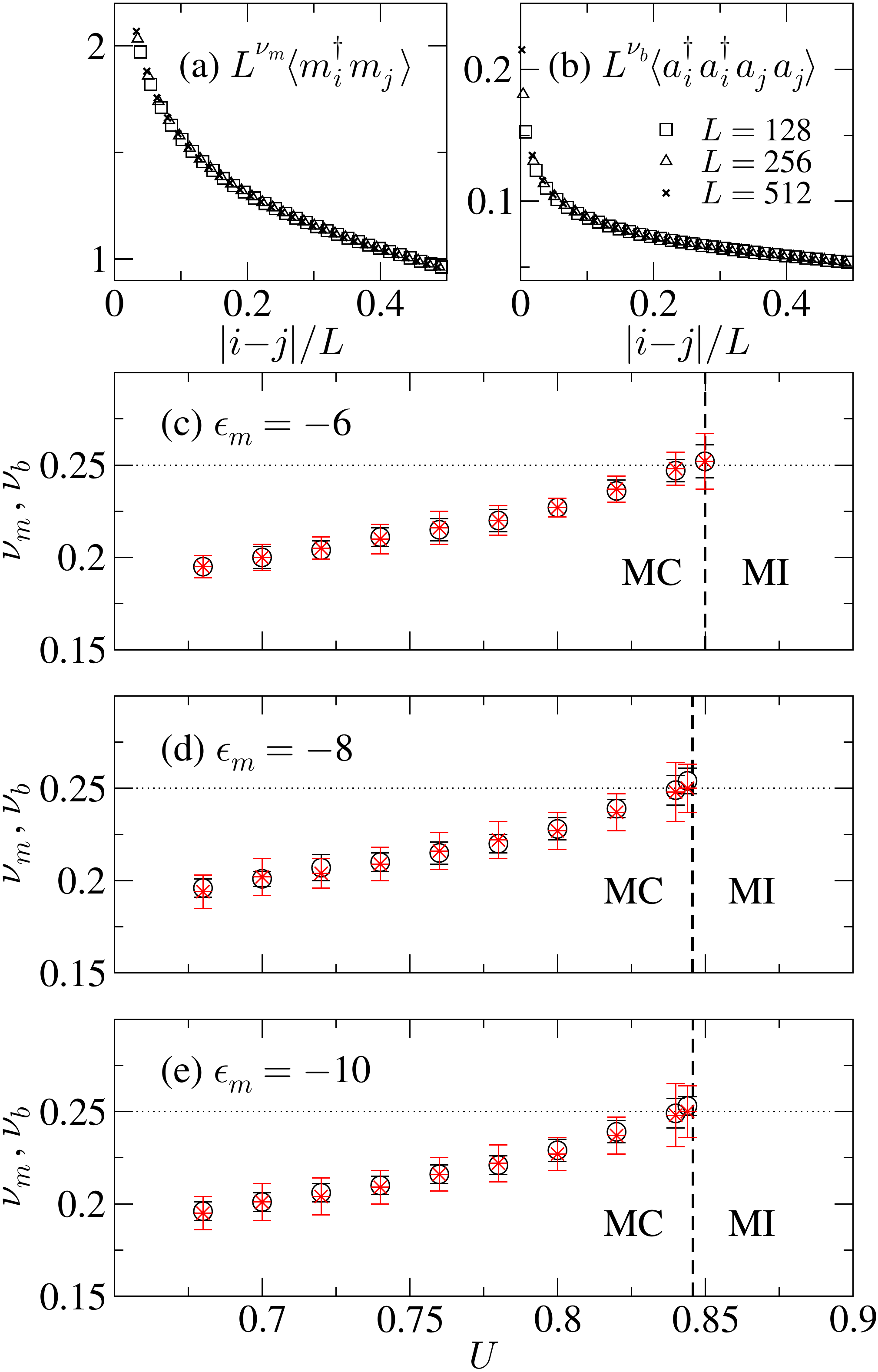}
\caption{(color online). DMRG results for the two-point functions of
  (a) molecules and (b) atomic bilinears in the MC phase with open
  boundaries and up to $L=512$.  We extract the molecular and bilinear
  exponents, $\nu_m$ and $\nu_b$, by finite-size scaling collapse of
  the data for different system sizes.  In (c), (d) and (e) we show
  the resulting evolution of $\nu_m$ (circles) and $\nu_b$ (stars),
  for vertical scans through Fig.~\ref{Fig:PD} with fixed values of
  $\epsilon_m$.  The vertical dashed lines correspond to the location
  of the MC to MI transition obtained from the gap data. The molecular
  exponent reaches the value of $\nu_m=1/4$ at the MC to MI
  transition. This corresponds to a molecular KT transition and is
  analogous to the fixed density transition at the tips of the Mott
  lobes in the single-band Bose--Hubbard model. The critical exponent
  $\nu_b$ associated with the power-law decay of the atomic bilinears
  $\langle a^\dagger_ia^\dagger_ia_ja_j\rangle$ (stars) coincides with
  $\nu_m$.}
\label{Fig:nummc}
\end{figure}

In order to explore these field theory predictions in more detail we
perform DMRG on the 1D Hamiltonian (\ref{atmolham}).  The predicted
behavior is well supported by our simulations in
Fig.~\ref{Fig:discorr}.
The molecules and atomic bilinears show power-law behavior with the
same exponent, $\nu_m=\nu_b$, whilst the atomic two-point function
shows exponential decay. Our DMRG results also indicate that this
behavior persists into the regime close to the Mott insulating phase
boundary shown in Fig.~\ref{Fig:PD}.  In particular, the molecular
correlation exponent reaches the value of $\nu_m=1/4$ at the MI
boundary; see Fig.~\ref{Fig:nummc}.
This is consistent with a molecular Kosterlitz--Thouless (KT)
\cite{KT:KT,Kosterlitz:Critical} transition. It is analogous to the
behavior at the tips of the Mott lobes in the single-band
Bose--Hubbard model \cite{Kuhner:1DBH,Giamarchi:Res,Buchler:Commens}
where the Luttinger liquid parameter takes the value
$K=1/(2\pi\nu)=2/\pi$ in the normalization conventions of
Eq.~(\ref{scalar}). The latter transition takes place at constant
density, and is therefore compatible with our canonical ensemble
constraint, $\rho_{\rm T}=2$.

We recall that in deriving the above correlation functions we have
neglected the coupling term, ${\mathcal L}_{\vartheta\phi}$ in the
low-energy Lagrangian so that the expressions factorize into
independent ${\rm U}(1)$ and ${\mathbb Z}_2$ contributions. The good
agreement with DMRG lends {\em a postiori} support to this
approximation within the explored region of the phase diagram.

\subsubsection{${\mathbb Z}_2$ Ordered AC+MC Phase}
\label{Sect:Ordgreen}

In the ${\mathbb Z}_2$ ordered phase the molecular Green's function
\be \langle m^\dagger(x)m(0)\rangle\propto \langle e^{-i\vartheta(x)}\
e^{i\vartheta(0)}\rangle\sim\left(\frac{a_0}{x}\right)^{\nu_m},
\label{mmord}
\ee
continues to decay as a power-law. In addition, the atomic Green's
function \be \langle a^\dagger(x)a(0)\rangle\sim\langle
\phi\rangle^2\langle e^{-i\frac{\vartheta(x)}{2}}\
e^{i\frac{\vartheta(0)}{2}}\rangle
\sim\left(\frac{a_0}{x}\right)^{\nu_a}
\label{aaord}
\ee also decays as a power-law, where the atomic correlation exponent
$\nu_a=\nu_m/4$ is locked to the molecular exponent by a factor of one
quarter \cite{Lee:Bosefeshbach}. This is a consequence of the Feshbach
coupling which ties the phases of the atomic and molecular condensates
together. Note that in writing Eq.~(\ref{aaord}), we approximate the
result for the two-point function of the Ising order parameter at
leading order \cite{Vaidya:Trans}: \be \langle
\phi(x)\phi(0)\rangle\sim \langle\phi\rangle^2\left[1+\pi^{-2}{\rm
    F}(x/\xi)\right]\approx \langle \phi\rangle^2, \label{phicorr}\ee
where \be {\rm F}(z)=z^2[{\rm K}_1^2(z)-{\rm K}_0^2(z)]-z{\rm
  K}_0(z){\rm K}_1(z) +\frac{1}{2}{\rm K}_0^2(z),
\label{Fdef}\ee and ${\rm K}_0(z)$ and ${\rm K}_1(z)$ are Bessel
functions.
\begin{figure}
\includegraphics[width=3.2in,clip=true]{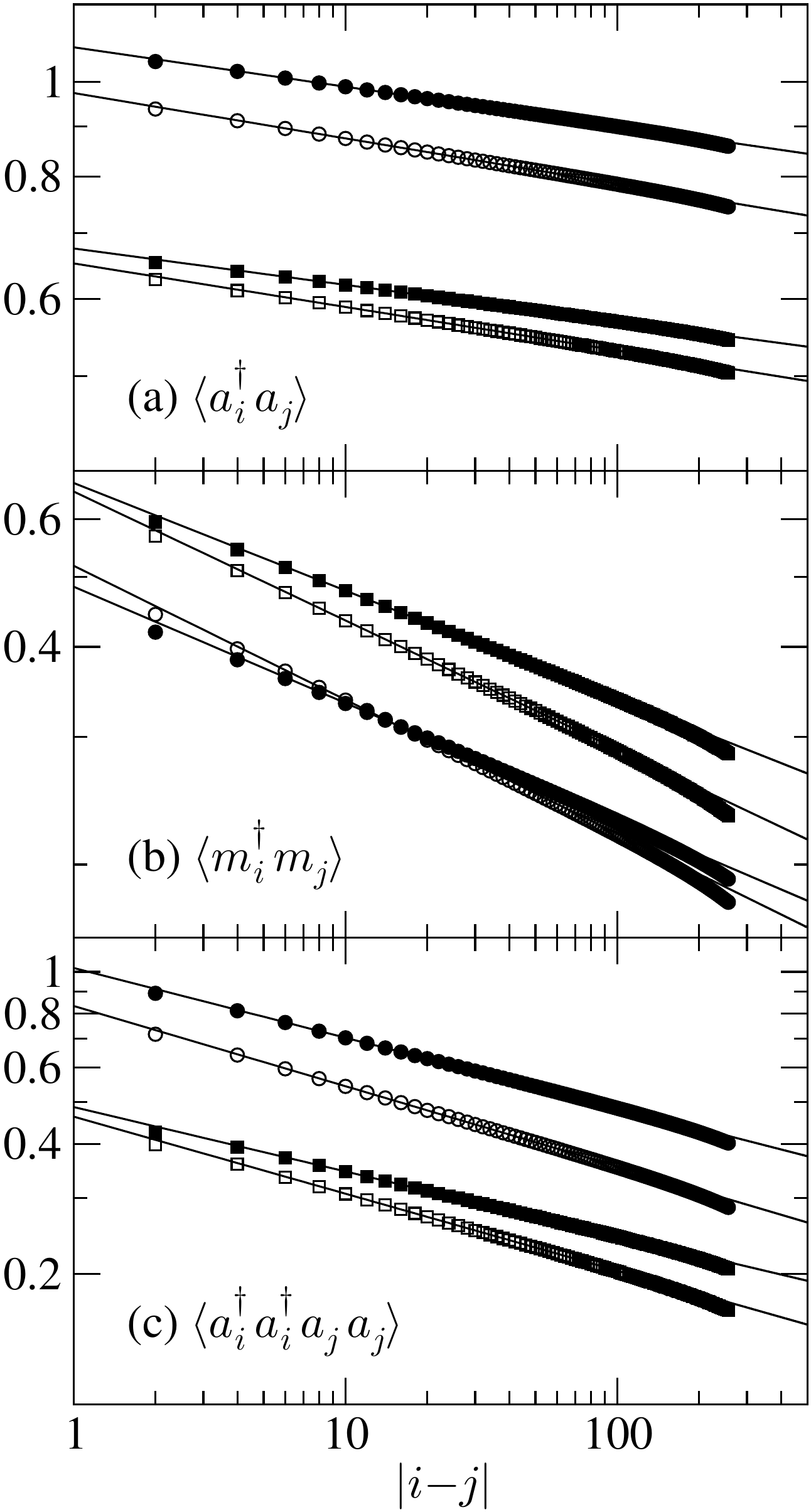}
\caption{Correlation functions in the ${\mathbb Z}_2$ ordered AC+MC
  phase obtained by DMRG on the 1D Hamiltonian (\ref{atmolham}) with
  $L=512$ and open boundaries. We use the same parameters as in
  Fig.~\ref{Fig:PD} and set $U=0.7$ (open), $U=0.5$ (filled),
  $\epsilon_m=-2$ (circles) and $\epsilon_m=-3$ (squares). (a) Atomic
  Green's function $\langle a^\dagger_ia_j\rangle$ showing power-law
  decay, in contrast to Fig.~\ref{Fig:discorr}(a). (b) Molecular
  Green's function $\langle m^\dagger_im_j\rangle$ showing power-law
  behavior. The exponent tracks the atomic exponent in (a) up to
  a factor of $4$; see Fig.~\ref{Fig:ACMCunity}. (c) Bilinears of
  atoms $\langle a^\dagger_ia^\dagger_ia_ja_j\rangle$ showing
  power-law behavior with the same exponent as the molecular Green's
  function in panel (b). This establishes the AC+MC phase as a pairing
  phase of atoms in the presence of atomic condensation.}
\label{Fig:ordcorr}
\end{figure}
These predictions of power-law behavior, as given by
Eqs.~(\ref{mmord}) and (\ref{aaord}), are well supported by our
numerical simulations as shown in Fig.~\ref{Fig:ordcorr}. The locking
of the atomic and molecular correlation exponents $\nu_a=\nu_m/4$ is
also observed.  In addition, these robust features persist into the
large-$U$ regime where field theory arguments are no longer strictly
valid. In particular, the atomic and molecular correlation functions
remain as power-laws right up to the MI boundary shown in
Fig.~\ref{Fig:PD}.
\begin{figure}
\includegraphics[width=3.2in,clip=true]{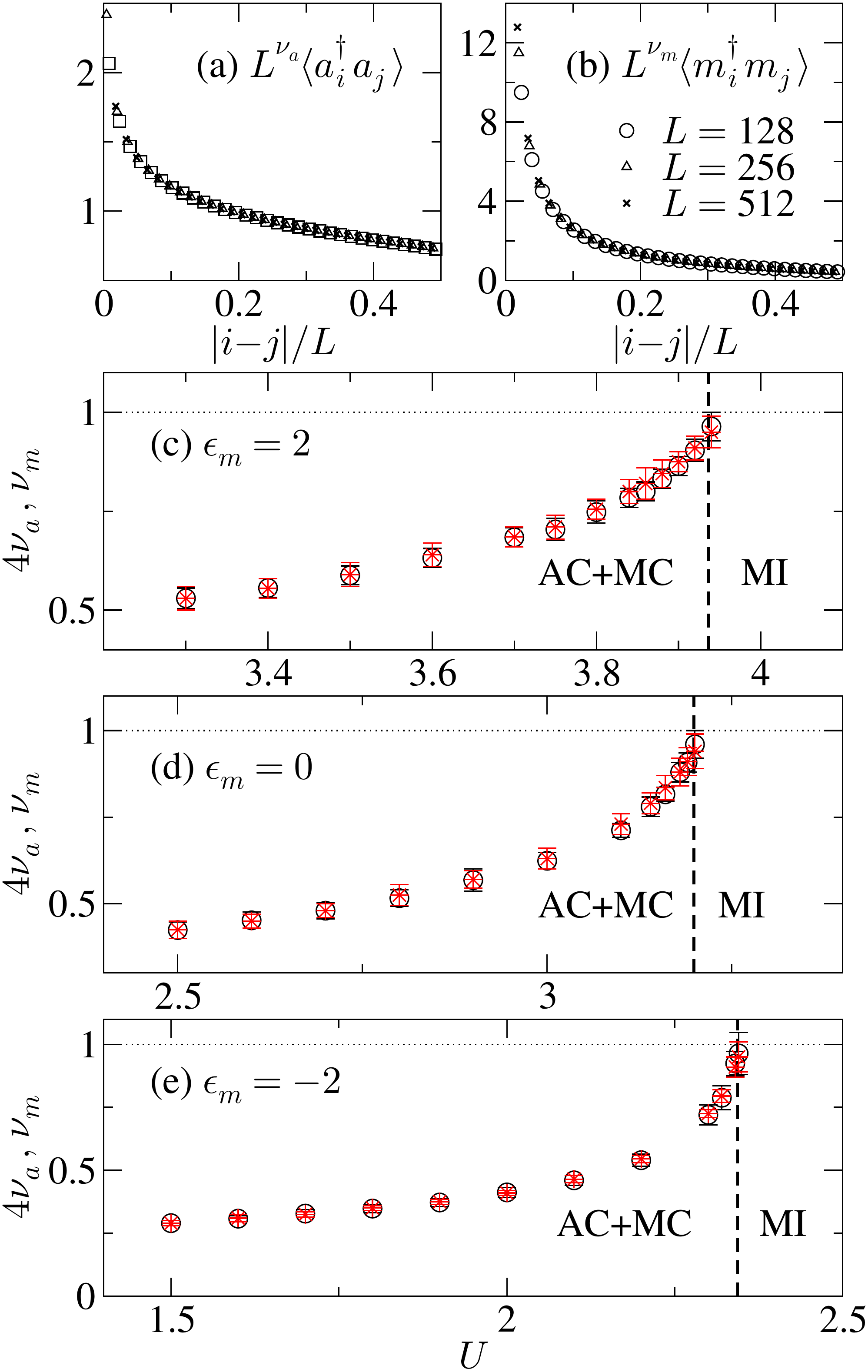}
\caption{(color online). DMRG results for the two-point functions of
  (a) atoms and (b) molecules in the AC+MC phase with open boundaries
  and up to $L=512$.  We extract the atomic and molecular exponents,
  $\nu_a$ and $\nu_m$, by finite-size scaling collapse of the data for
  different system sizes.  In (c), (d) and (e) we show the resulting
  evolution of $4\nu_a$ (circles) and $\nu_m$ (stars), for vertical
  scans through Fig.~\ref{Fig:PD} with fixed values of $\epsilon_m$.
  The vertical dashed lines correspond to the location of the AC+MC to
  MI transition obtained from the gap data. The data confirm the
  locking of the atomic and molecular exponents via the relation
  $\nu_m=4\nu_a$.  The exponents reach the values of $\nu_a=1/4$ and
  $\nu_m=1$ at the MI boundary. This is consistent with an atomic KT
  transition. It is analogous to the behavior at the tips of the Mott
  lobes in the single-band Bose--Hubbard model.}
\label{Fig:ACMCunity}
\end{figure}
We find that the atomic exponent $\nu_a$ reaches the value of
$\nu_a=1/4$ at the AC+MC to MI transition; see
Fig.~\ref{Fig:ACMCunity}. This is consistent with an atomic KT
transition, as occurs at the tips of the Mott lobes in the single-band
Bose--Hubbard model. At the same time, the molecular exponent $\nu_m$
reaches the value of $\nu_m=1$ due to the aforementioned exponent
locking; see Fig.~\ref{Fig:ACMCunity}.  The presence of this molecular
superfluid close to the MI boundary, clearly supports the absence of a
single component atomic superfluid phase in this 1D setting, in
contrast to the findings of Ref.~\cite{Rousseau:Fesh}. This is also
compatible with mean field theory in higher dimensions
\cite{Rad:Atmol,Romans:QPT,Radzi:Resonant} where atomic condensation
is always accompanied by molecular condensation due to the structure
of the Feshbach term, $H_{\rm F}$.  We will return to this issue in
Sec.~\ref{Sect:Mom} in our discussion of the corresponding
zero-momentum occupation numbers and the visibility.

\subsubsection{Mixed Correlation Functions}
\begin{figure}
\includegraphics[width=3.2in,clip=true]{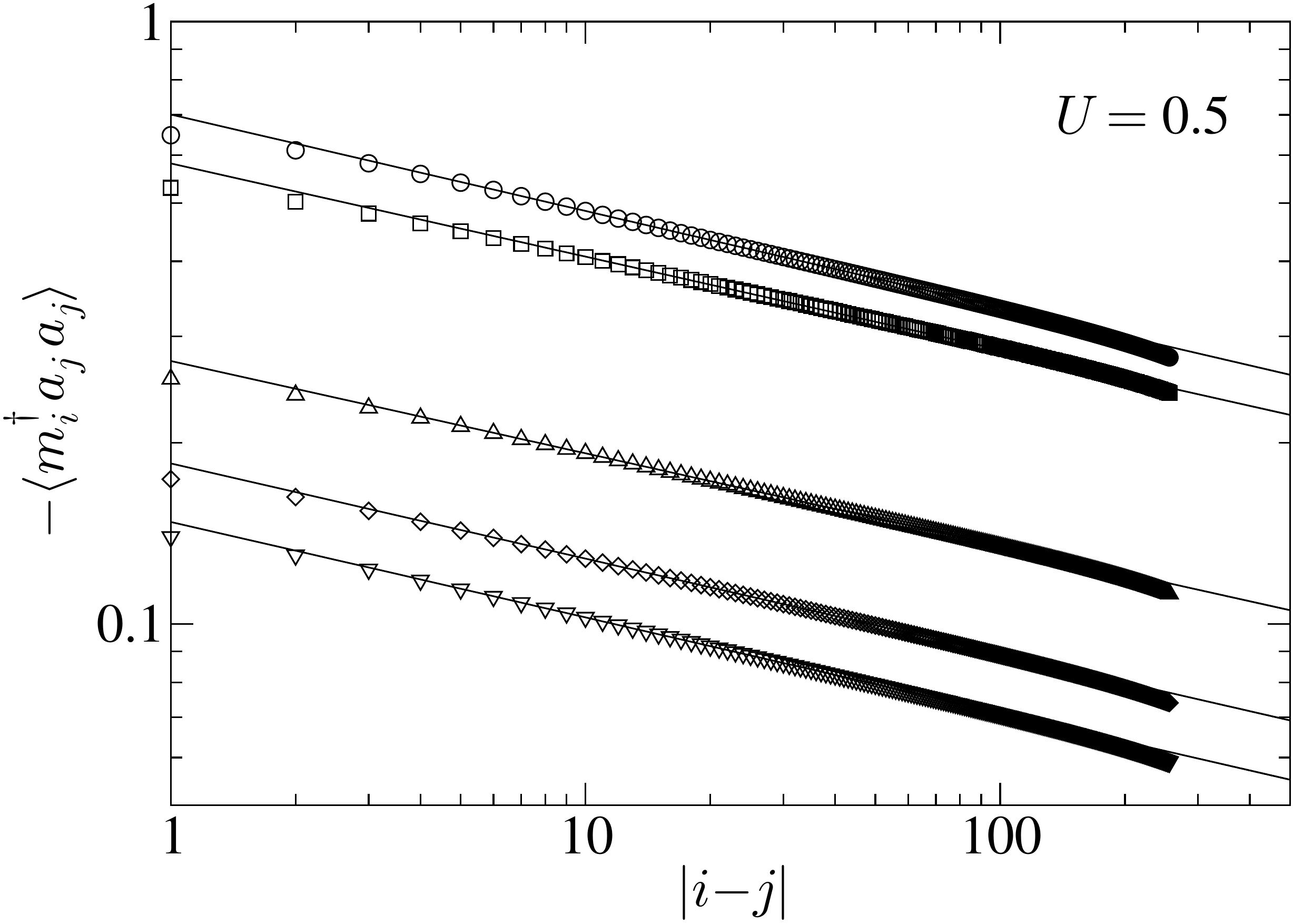}
\caption{DMRG results for the mixed correlation function
  $-\langle m^\dagger_ia_ja_j\rangle$ with $L=512$ sites,
  open boundaries and $U=0.5$. The data correspond to $\epsilon_m=-2$
  (circles), $\epsilon_m=-3$ (squares), $\epsilon_m=-4$ (up
  triangles), $\epsilon_m=-5$ (diamonds), $\epsilon_m=-6$ (down
  triangles) and show power-law behavior in both the AC and AC+MC
  phases. As predicted by Eq.~(\ref{maacorr}), the exponents agree
  with those of the molecular Green's function in panels (b) of
  Figs.~\ref{Fig:discorr} and \ref{Fig:ordcorr}.}
\label{Fig:mixed}
\end{figure}

In addition to the purely atomic or molecular Green's functions, it is
also instructive to examine the mixed correlation functions involving
both atoms and molecules. It follows from the decomposition
(\ref{amising}) that \be \langle m^\dagger(x)a(0)a(0)\rangle \sim
\sqrt{\rho_m}\,\langle\phi^2\rangle\left(\frac{a_0}{x}\right)^{\nu_m},
\label{maacorr}
\ee decays as a power-law with the same exponent as the molecular
Green's function.  Once again, this reflects the phase locking of the
atomic and molecular condensates due to the Feshbach term, and is
present in both the MC and AC+MC phases. This behavior is in very good
agreement with our DMRG simulations as shown in
Fig.~\ref{Fig:mixed}. In particular, the power-law exponent tracks
those displayed in Figs.~\ref{Fig:discorr}(b) and \ref{Fig:ordcorr}(b)
for $\langle m^\dagger(x)m(0)\rangle$.

\subsection{Density Correlation Functions}
\begin{figure}
\includegraphics[width=3.2in,clip=true]{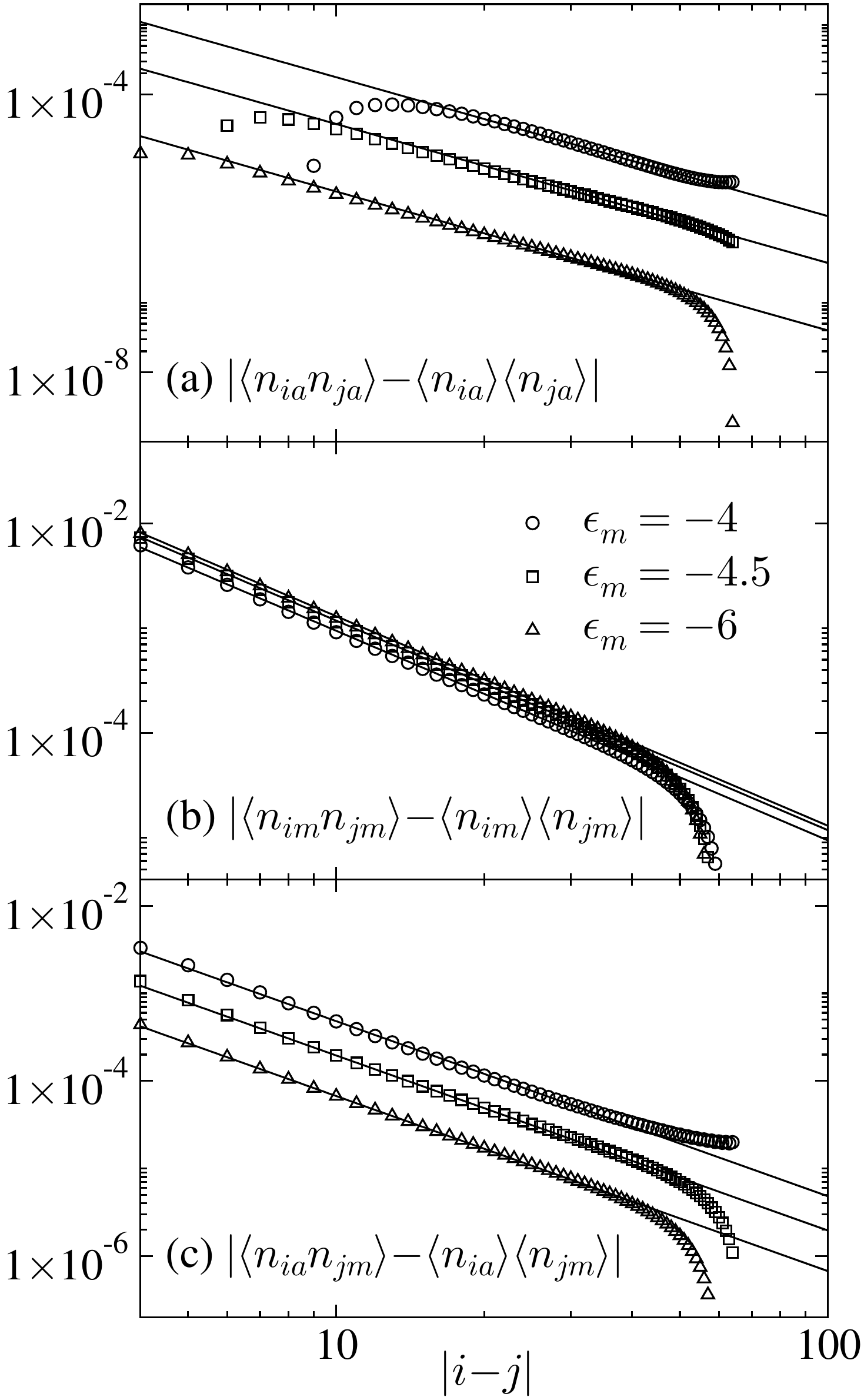}
\caption{DMRG results for the connected density correlation functions
  in the ${\mathbb Z}_2$ disordered MC phase for the parameters used
  in Fig.~\ref{Fig:PD}.  The values of $\epsilon_m$ are indicated in
  panel (b).  We use open boundaries with $L=128$ and set $U=0.7$.
  (a) $|\langle n_{ia} n_{ja}\rangle-\langle n_{ia}\rangle \langle
  n_{ja}\rangle|$. (b) $|\langle n_{im} n_{jm}\rangle-\langle
  n_{im}\rangle \langle n_{jm}\rangle|$. (c) $|\langle
  n_{ia}n_{jm}\rangle-\langle n_{ia}\rangle\langle
  n_{jm}\rangle|$. The results are in agreement with the leading
  $1/x^2$ dependence predicted by Eq.~(\ref{densdens}).}
\label{Fig:densdis}
\end{figure}

\begin{figure}
\includegraphics[width=3.2in,clip=true]{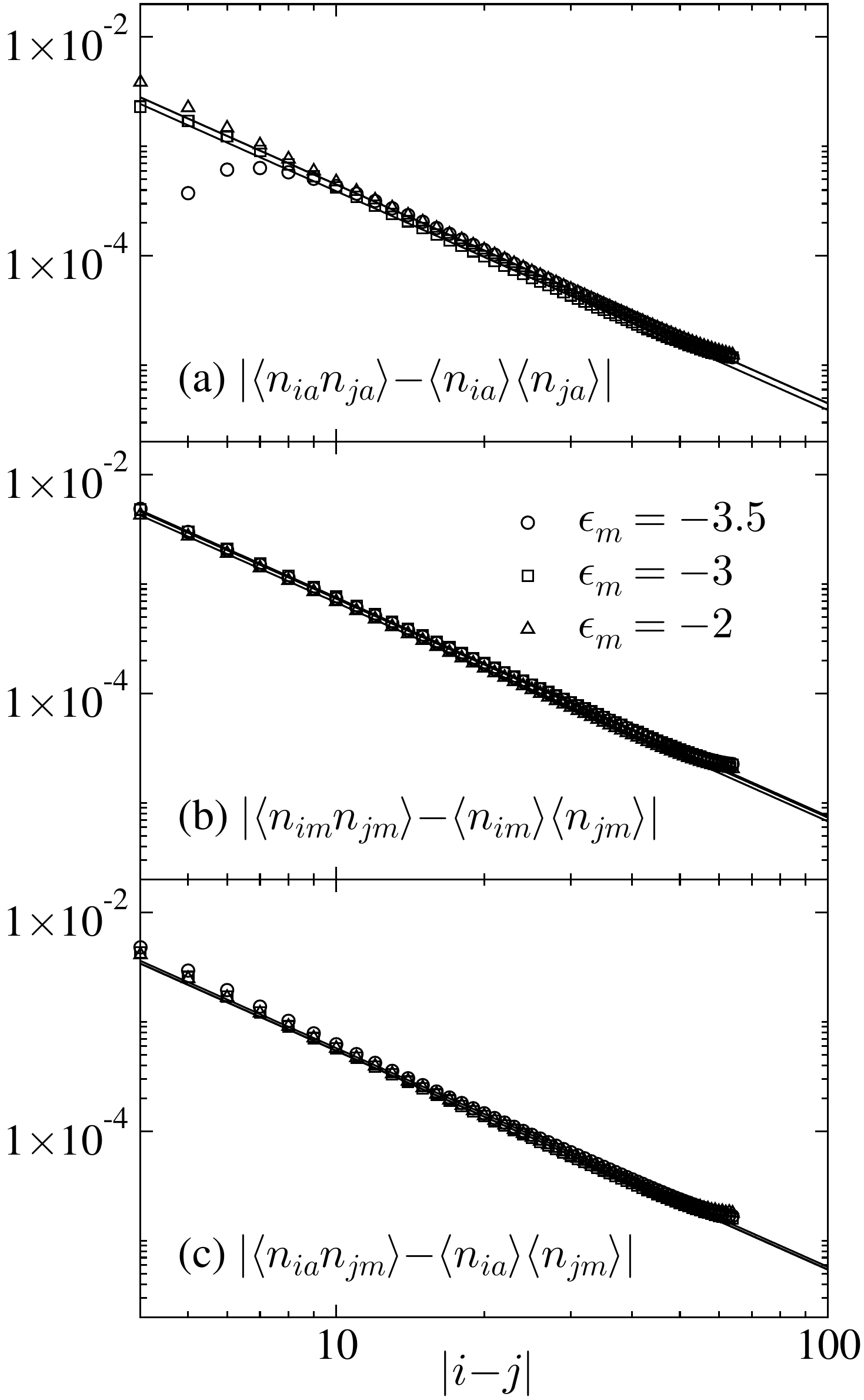}
\caption{DMRG results for the connected density correlation functions
  in the ${\mathbb Z}_2$ ordered AC+MC phase for the parameters used
  in Fig.~\ref{Fig:PD}. The values of $\epsilon_m$ are indicated in
  panel (b). We use open boundaries with $L=128$ and set
  $U=0.7$. (a) $|\langle n_{ia} n_{ja}\rangle-\langle n_{ia}\rangle
  \langle n_{ja}\rangle|$. (b) $|\langle n_{im}n_{jm}\rangle-\langle
  n_{im}\rangle \langle n_{jm}\rangle|$. (c) $|\langle
  n_{ia}n_{jm}\rangle-\langle n_{ia}\rangle \langle
  n_{jm}\rangle|$. The results are in agreement with the leading
  $1/x^2$ dependence predicted by Eq.~(\ref{densdens}).}
\label{Fig:densord}
\end{figure}

Having discussed the atomic and molecular Green's functions we now
turn our attention to the correlation functions of the local
densities. Denoting $n_m(x)\equiv m^\dagger(x)m(x)$ and $n_a(x)\equiv
a^\dagger(x)a(x)$ one obtains \bea
n_m(x)&\sim&\rho_m+\gamma_1\partial_x\vartheta+\ldots,\nn
n_a(x)&\sim&\rho_a+\gamma_2\partial_x\vartheta
+\gamma_3:\phi^2(x):+\ldots,
\label{densexp}
\eea where $\rho_m$ and $\rho_a$ are the average molecular and atomic
densities, and $\gamma_1$, $\gamma_2$, $\gamma_3$ are constants. Here
we use the primary correspondence given in Eq.~(\ref{amising}), and
combine the exponentials by point-splitting and the short distance
operator product expansion. The expansion (\ref{densexp}) incorporates
the effects of density fluctuations and it follows that the
density-density correlations have the same leading dependence in both
the MC and AC+MC phases:
\begin{equation}
\langle n_\alpha(x)n_\beta(0)\rangle \simeq \rho_\alpha\rho_\beta
+\frac{C_{\alpha\beta}}{x^2}+\dots,
\label{densdens}
\end{equation} 
where $\alpha,\beta\in a,m$ and $C_{\alpha\beta}$ are non-universal
constants.  This is confirmed by our DMRG results in
Figs.~\ref{Fig:densdis} and \ref{Fig:densord}.

\section{Momentum Space Observables}
\label{Sect:Mom}
\begin{figure}
\includegraphics[width=3.2in]{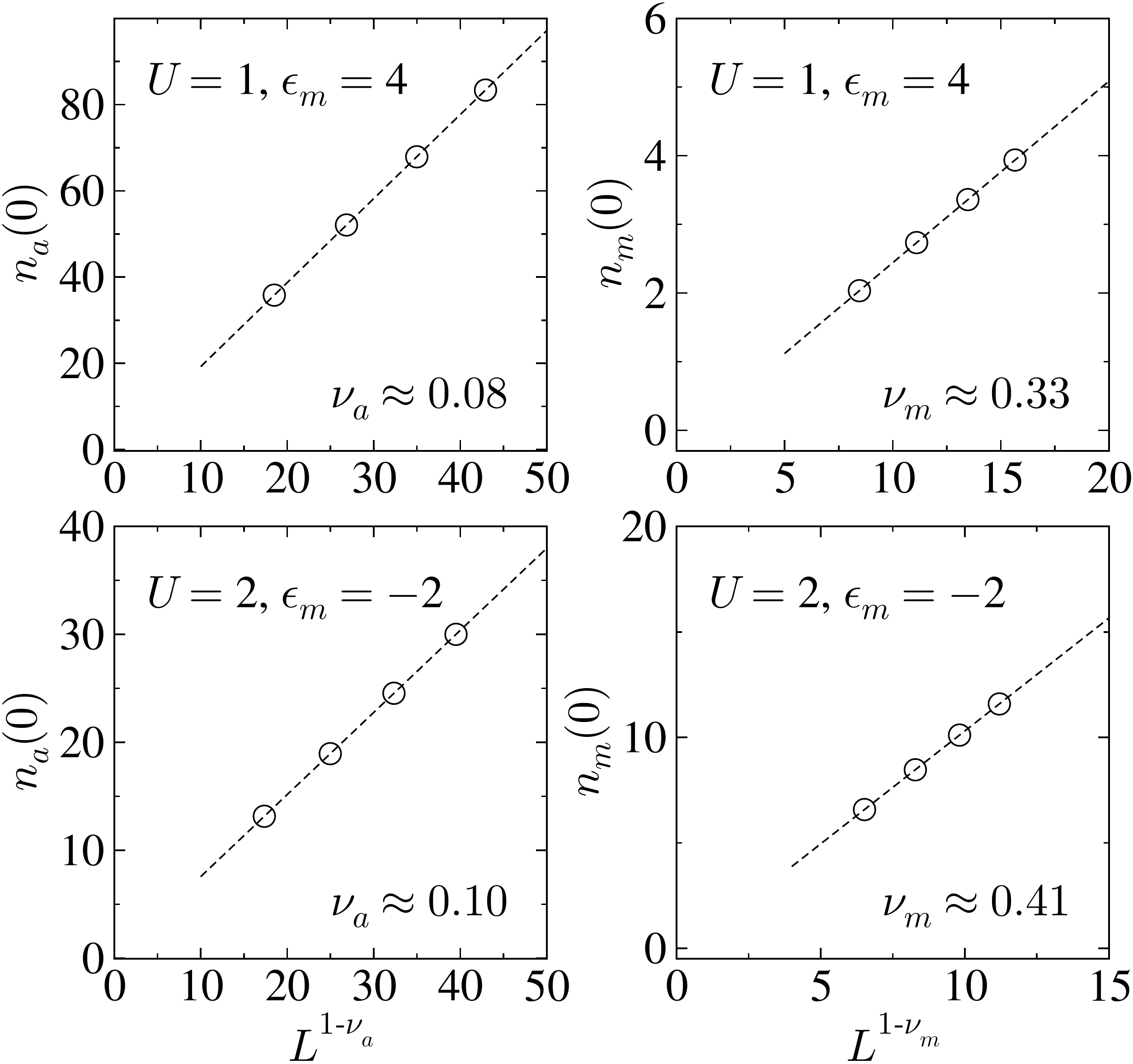}
\caption{DMRG results for the dependence of the zero-momentum
  occupation numbers $n_a(0)$ and $n_m(0)$ on system size $L$ within
  the AC+MC phase shown in Fig.~\ref{Fig:PD}.  The results are
  consistent with algebraic correlations for both atoms and molecules
  with locked exponents $\nu_m=4\nu_a$.  The presence of molecular
  superfluidity in the lower panels confirms the absence of an AC
  phase close to the MI boundary.}
\label{Fig:Zmomdiv}
\end{figure}

In the previous section we have focused directly on the superfluid
correlation functions due to the absence of continuous symmetry
breaking in 1D.  However, a useful diagnostic of superfluidity in
higher dimensions is the divergence of the occupation number
\begin{equation}
n_\alpha(k)= \frac{1}{L}\sum_{i,j=1}^L e^{ik(i-j)}
\langle a_{\alpha,i}^\dagger a_{\alpha,j}^{\phantom\dagger}\rangle
\label{nkdef}
\end{equation}
at zero momentum, $k=0$. This quantity was recently used in
Ref.~\cite{Rousseau:Fesh}, in conjunction with visibility data, to
argue in favor of a single component atomic superfluid phase in the 1D
system (\ref{atmolham}).  In view of our results in the previous
sections, which show the presence of both atomic {\em and} molecular
superfluidity right up to the Mott boundary in Fig.~\ref{Fig:PD}, we
revisit this issue here.  As shown in Fig.~\ref{Fig:ACMCunity}, the
atomic and molecular correlation functions in the AC+MC phase are
power-laws, $\langle m^\dagger(x)m(0)\rangle\sim x^{-\nu_m}$ and
$\langle a^\dagger(x)a(0)\rangle\sim x^{-\nu_m/4}$, with locked
exponents. Substituting these asymptotic forms into Eq.~(\ref{nkdef})
suggests that the zero-momentum occupation numbers depend on system
size according to \cite{Cha:Momdist,Cha:Momdist2}
\begin{equation}
n_m(0)\sim {\mathcal A}_m+{\mathcal B}_mL^{1-\nu_m},\quad n_a(0)\sim {\mathcal A}_a+{\mathcal B}_a L^{1-\nu_m/4},
\label{nmal}
\end{equation} 
where ${\mathcal A}_{a,m}$ and ${\mathcal B}_{a,m}$ are constants. 
In particular, since the molecular exponent, $\nu_m$, only reaches
unity at the Mott phase boundary (see Figs. \ref{Fig:PD} and
\ref{Fig:ACMCunity}) both of these zero-momentum occupation numbers 
are expected to diverge with increasing system size. 
This is supported by our DMRG results as 
shown in Fig.~\ref{Fig:Zmomdiv}. 
However, it is evident from Eq.~(\ref{nmal}) that $n_m(0)$ diverges
very slowly with increasing system size close to the MI boundary since
$\nu_m\rightarrow 1$.  In the absence of a detailed finite-size
scaling analysis this may lead to the erroneous conclusion of a purely
atomic superfluid. In addition, our findings suggest the absence of
any change in behavior in the convergence properties of $n_m(0)$ as
$L\rightarrow\infty$, which could be misinterpreted as a quantum phase
transition to a purely atomic superfluid. In general, in this 1D
setting, the zero-momentum occupation number is a poor diagnostic of
superfluid transitions, since it may simply reflect a change in the
{\em value} of the critical exponent within a superfluid phase, rather
than the onset of exponential correlations. Direct evaluation of the
correlation functions $\langle a^\dagger(x)a(0)\rangle$ and $\langle
m^\dagger(x)m(0)\rangle$ provides a clearer picture in 1D, especially
in the case of a finite size system.  Our results are fully consistent
with the absence of a purely atomic superfluid phase in this region of
the phase diagram.  This is compatible with the predictions of mean
field theory in higher dimensions
\cite{Rad:Atmol,Romans:QPT,Radzi:Resonant}.

\begin{figure}
\includegraphics[width=3.2in,clip=true]{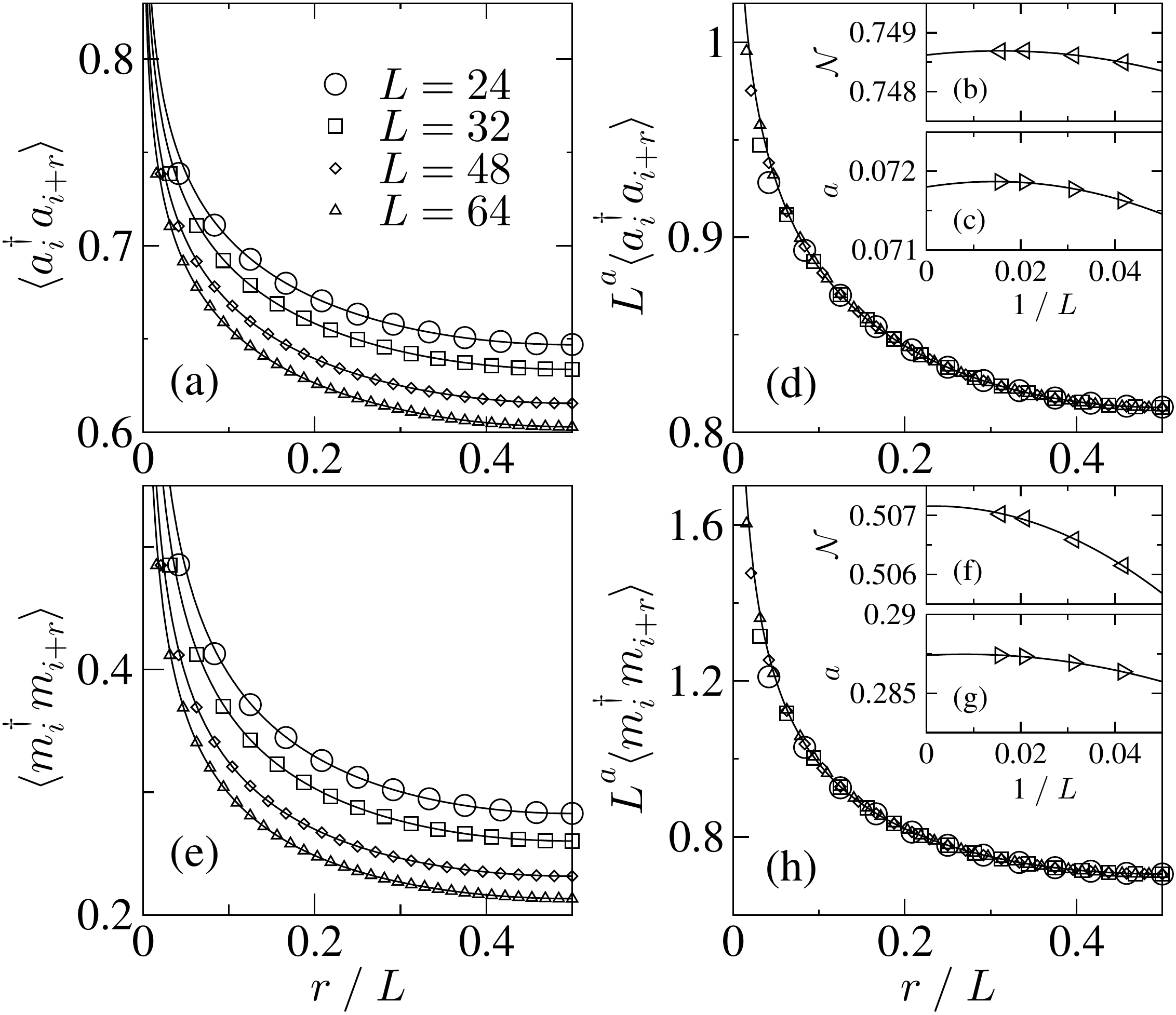}
\caption{Atomic and molecular correlation functions within the AC+MC
  phase obtained by DMRG with periodic boundary conditions. We set
  $\epsilon_m=-2$ and $U=1.5$.  (a) Atomic correlation function
  $\langle a_i^\dagger a_{i+r}\rangle$ as a function of the reduced
  separation $r/L$ for different system sizes. (b) Normalization
  factor ${\mathcal N}$ obtained from panel (a) using
  Eq.~(\ref{conftranscorr}). (c) Correlation exponent $a$ obtained
  from panel (a) using Eq.~(\ref{conftranscorr}). (d) Rescaling the
  data in panel (a) using the extracted exponent $a$ leads to data
  collapse.  This confirms the applicability of the conformal result
  (\ref{conftranscorr}) within the AC+MC phase. This corresponds to
  power-law atomic correlations for separations $r\gtrsim 3a_0$. The
  remaining panels show the corresponding results for molecules.}
\label{Fig:Corrcollapse}
\end{figure}

\begin{figure}
\includegraphics[width=3.2in,clip=true]{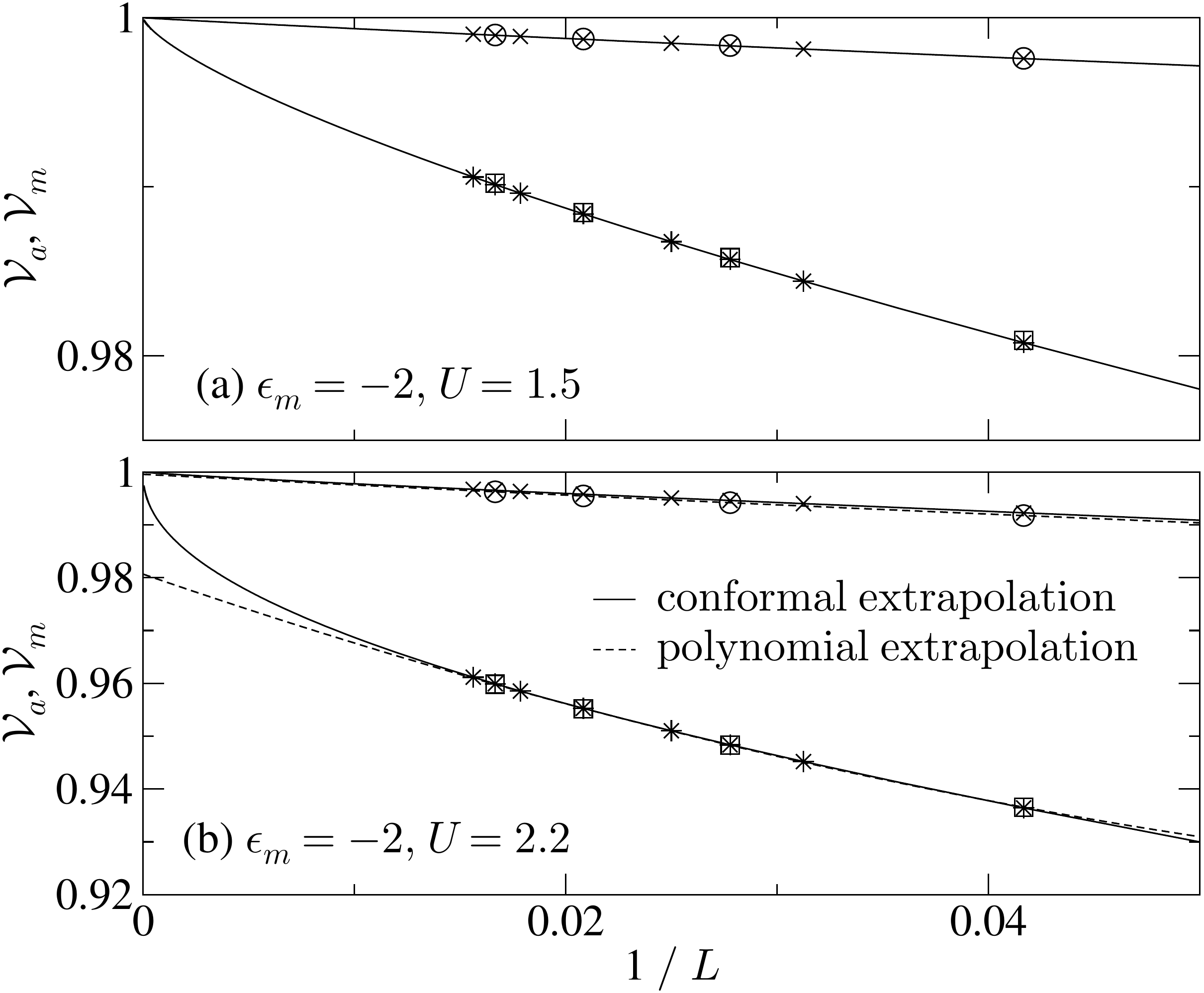}
\caption{Finite-size scaling of the atomic and molecular visibilities
  within the AC+MC phase.  The circles correspond to ${\mathcal V}_a$
  and the squares to ${\mathcal V}_m$ obtained by Fourier
  transformation of the correlation functions obtained by DMRG.  The
  crosses and stars correspond to Fourier transformation of the
  conformal result (\ref{conftranscorr}) supplemented by exact DMRG
  results for the correlators at small separations $r\le 3a_0$.  The
  solid line indicates the results of conformal extrapolation
  (described in the text and justified by the scaling collapse
    in Fig.~\ref{Fig:Corrcollapse}) supplemented by the exact DMRG
  results for small separations $r\le 3a_0$. (a) With $\epsilon_m=-2$
  and $U=1.5$ both ${\mathcal V}_a$ and ${\mathcal V}_m$ extrapolate
  to unity in the thermodynamic limit. (b) Close to the MI transition
  with $\epsilon_m=-2$ and $U=2.2$ both ${\mathcal V}_a$ and
  ${\mathcal V}_m$ approach unity as $L\rightarrow\infty$. This is in
  direct contrast to naive polynomial extrapolation (dashed) which
  erroneously suggests that the molecular visibility is less than
  unity.}
\label{Fig:Corrextrap}
\end{figure}

\begin{figure}
\includegraphics[width=3.2in,clip=true]{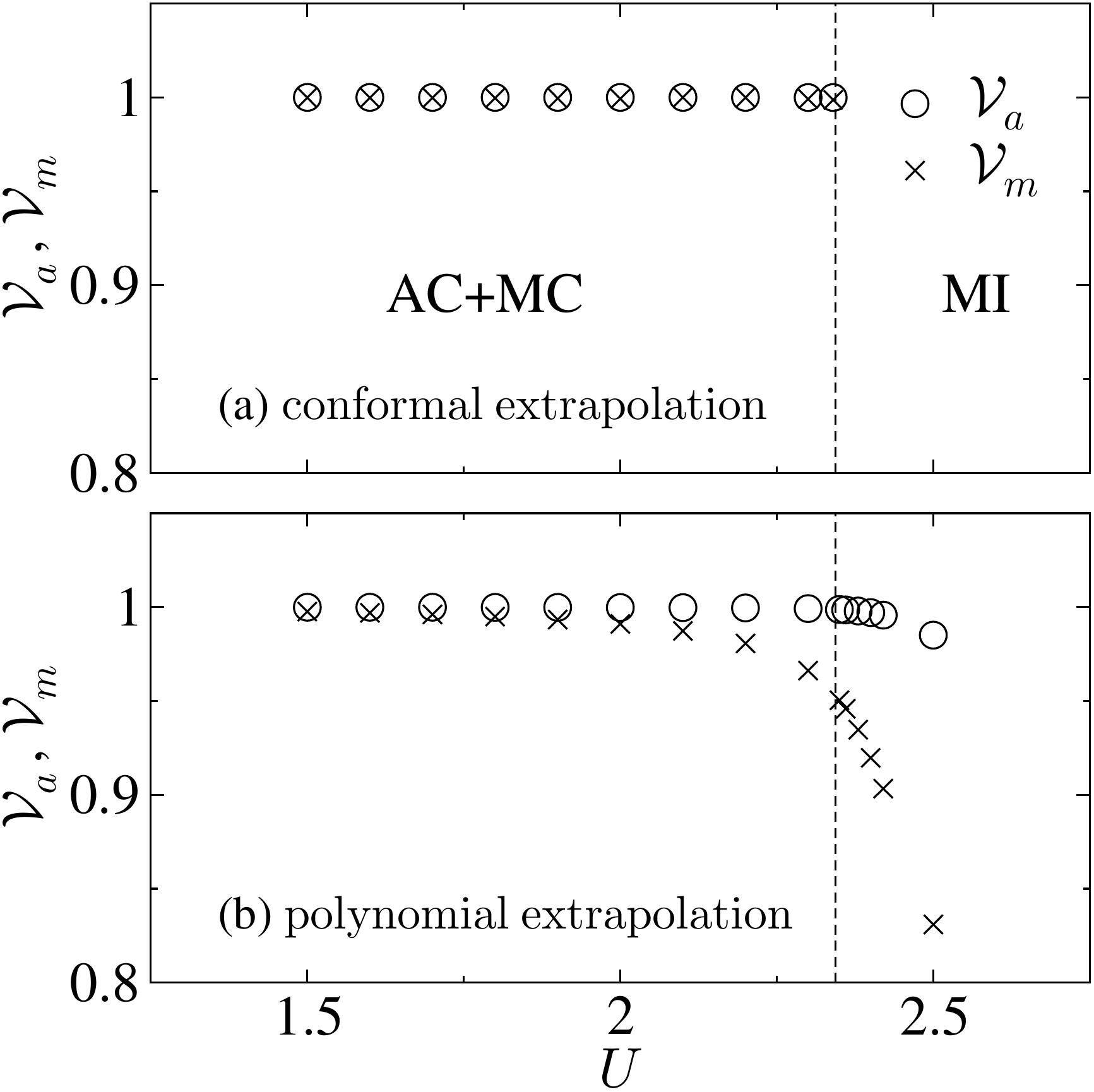}
\caption{(a) Atomic and molecular visibilities ${\mathcal V}_a$
  (circles) and ${\mathcal V}_m$ (crosses) within the AC+MC phase for
  $\epsilon_m=-2$ obtained by DMRG with up to $L=64$ and periodic
  boundaries.  We use the conformal extrapolation procedure described
  in the text in order to obtain the asymptotic results as
  $L\rightarrow\infty$. Both ${\mathcal V}_a$ and ${\mathcal V}_m$ are
  unity right up to the MI boundary, indicating the presence of both
  atomic and molecular superfluidity.  (b) Naive polynomial
  extrapolation erroneously suggests that the molecular visibility is
  less than unity in the AC+MC phase.}
\label{Fig:Vis}
\end{figure}

In addition to the zero-momentum occupation numbers, the authors of
Ref.~\cite{Rousseau:Fesh} also consider the visibility.  The
visibility is related to the momentum occupation numbers (\ref{nkdef})
via \cite{Gerbier:PC}
\begin{equation}
{\mathcal V}_\alpha\equiv \frac{n_\alpha^{\rm max}(k)-n_\alpha^{\rm min}(k)}{n_\alpha^{\rm max}(k)+n_\alpha^{\rm min}(k)},
\label{visdef}
\end{equation} 
where $n^{\rm max}$ ($n^{\rm min}$) is the maximum (minimum) in the
momentum space occupation number distribution. In the present context
this is identified as
\begin{equation}
{\mathcal V}_\alpha=\frac{n_\alpha(0)-n_\alpha(\pi)}{n_\alpha(0)+n_\alpha(\pi)}.
\label{viszeropi}
\end{equation}
In a superfluid phase where $n_\alpha(0)$ diverges with increasing
system size, the visibility ${\mathcal V}_\alpha$ approaches unity as
$L\rightarrow\infty$. In Ref.~\cite{Rousseau:Fesh} it was argued that
the molecular visibility within the AC+MC phase failed to saturate at
this value close to the MI boundary. In order to gain a quantitative
handle on this issue we need to exploit the finite-size dependence of
the superfluid correlations within the AC+MC phase. In a system with
periodic boundary conditions the two-point function of a primary field
${\mathcal O}(r)$ at position $r$ can be obtained by conformal
transformation \cite{Cardy:scal}:
\begin{equation}
  \langle {\mathcal O}(r_1){\mathcal O}(r_2)\rangle_L={\mathcal N}\left[\frac{\pi}{L\sin(\frac{\pi r}{L})}\right]^a,
\label{conftranscorr}
\end{equation}
where $a$ is the critical exponent in the thermodynamic limit,
$r=|r_1-r_2|$ is the separation, and ${\mathcal N}$ is a constant
pre-factor. It follows that the rescaled combination $L^a\langle
{\mathcal O}(r_1){\mathcal O}(r_2)\rangle_L$ is a prescribed scaling
function of the reduced separation $r/L$.  The confirmation of this
behavior for the atomic and molecular correlation functions within the
AC+MC phase is shown in Fig.~\ref{Fig:Corrcollapse}.
Given this agreement we may substitute the conformal result
(\ref{conftranscorr}) into Eq.~(\ref{nkdef}) in order to obtain formal
expressions for the finite-size dependence of the atomic and molecular
visibilities in Eq.~(\ref{visdef}).  In Fig.~\ref{Fig:Corrextrap} we
show the results of this conformal extrapolation, where we further
incorporate the exact DMRG results for the short distance behavior
with $r\le 3a_0$, where $a_0$ is the lattice spacing.
It is readily seen from the solid lines in Fig.~\ref{Fig:Corrextrap}
that both the atomic and molecular visibilities extrapolate to unity
in the thermodynamic limit. In particular, close to the MI boundary
there are strong deviations from the results that would be obtained by
naive polynomial extrapolation as indicated by the dashed lines.  In
Fig.~\ref{Fig:Vis}(a)
we use the conformal extrapolation procedure to track the atomic and
molecular visibilities within the AC+MC phase.  The results are
consistent with unity right up to the MI boundary.  For comparison, in
Fig.~\ref{Fig:Vis}(b) we show the results that would be inferred using
a naive polynomial extrapolation. The results are in accordance with
those of Ref.~\cite{Rousseau:Fesh}, but differ markedly from the
asymptotic visibilities obtained by conformal extrapolation as shown
in Fig.~\ref{Fig:Vis}(a).

To summarize the results of this section, within the AC+MC phase the
finite-size dependence of the atomic and molecular momentum space
diagnostics is in complete agreement with power-law correlations for
both the atoms and the molecules. This behavior persists right up to
the MI boundary and provides further evidence for the absence of a
purely AC phase. This is analogous to expectations in higher
dimensions arising from mean-field theory
analyses \cite{Rad:Atmol,Romans:QPT,Radzi:Resonant}.

\section{Entanglement Entropy}
\label{Sect:EE}
\begin{figure}
\includegraphics[width=3.2in,clip=true]{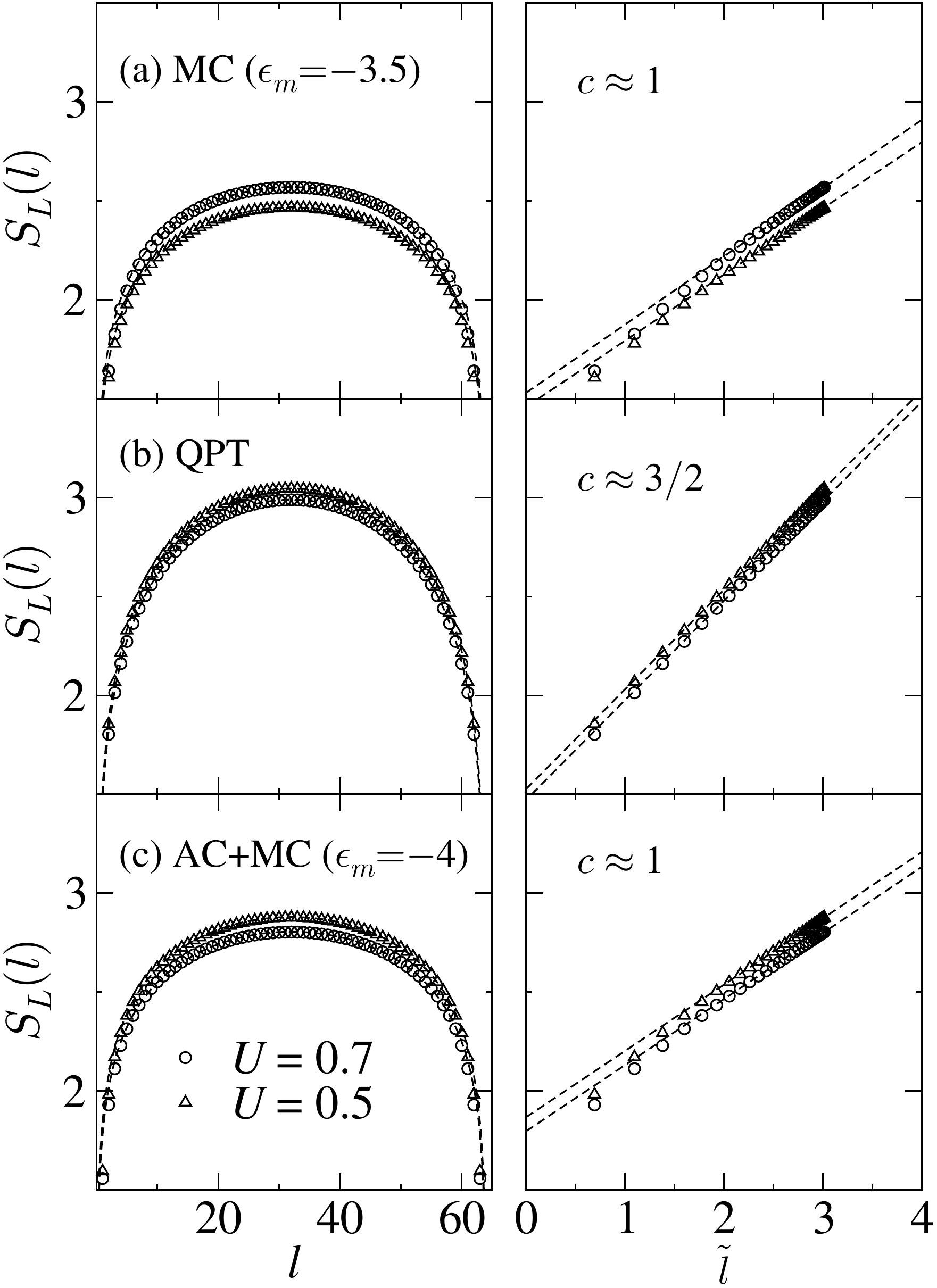}
\caption{Entanglement entropy $S_L(l)$ obtained by DMRG with $L=64$
  and periodic boundaries.  We consider horizontal scans through
  Fig.~\ref{Fig:PD} with $U=0.5$ and $U=0.7$.  (a) Within the MC phase
  with $\epsilon_m=-3.5$ we find $c\approx 1$ corresponding to a
  gapless superfluid.  (b) In the vicinity of the MC to AC+MC quantum
  phase transition we find $c\approx 3/2$. This corresponds to the
  presence of additional gapless Ising degrees of freedom coexisting
  with superfluidity.  (c) Within the AC+MC phase with $\epsilon_m=-4$
  we find $c\approx 1$ corresponding to an effective free boson. The
  panels on the right correspond to the same data as on the left, but
  are plotted against the conformal distance $\tilde l\equiv
  \ln[(L/\pi)\sin(\pi l/L)]$ in order to yield a linear plot with
  slope $c/3$. The offset between the different curves within each
  panel is due to the non-universal contribution in Eq.~(\ref{ee}).}
\label{Fig:Entang}
\end{figure}

\begin{figure}
\includegraphics[width=3.2in,clip]{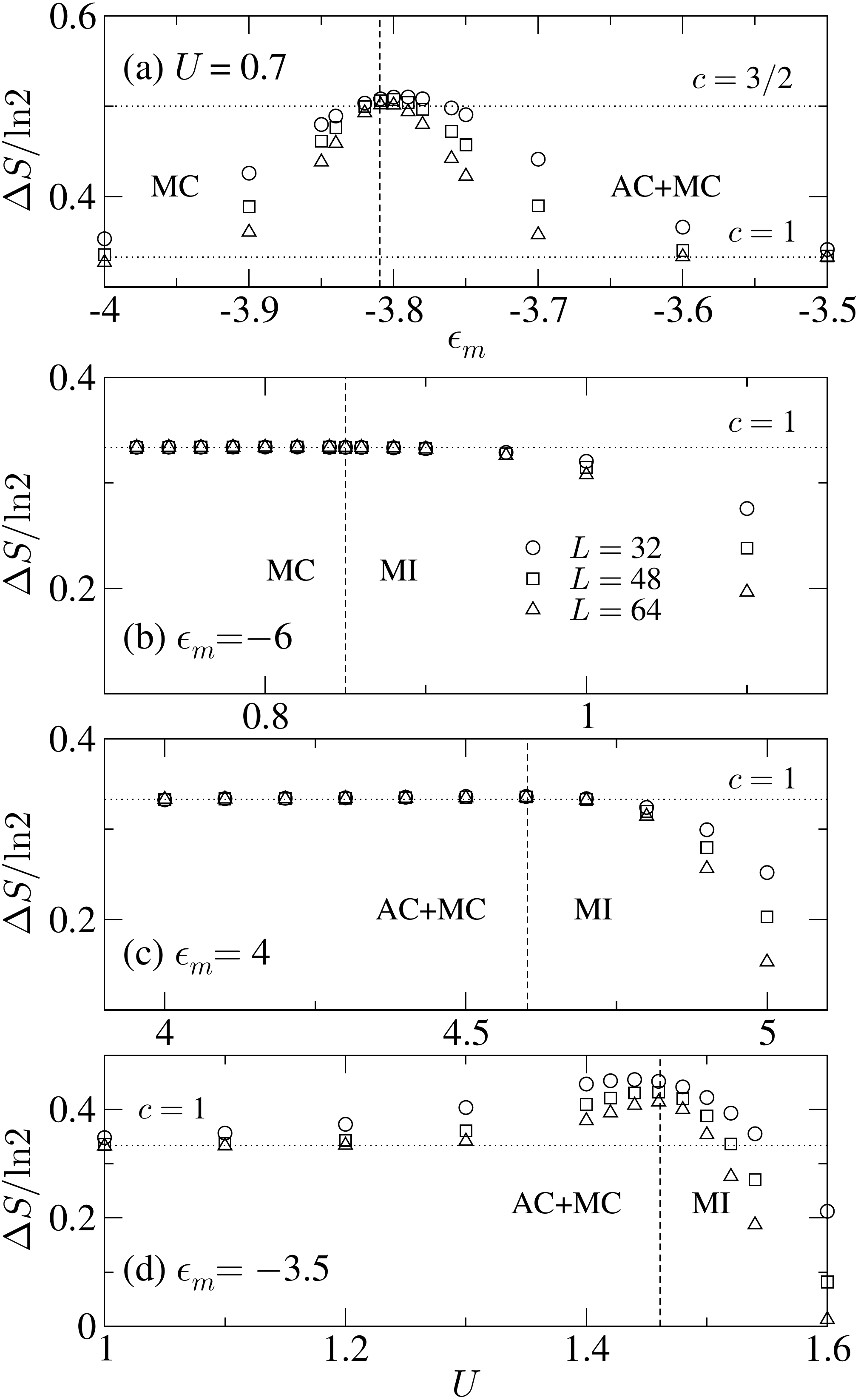}
\caption{Entanglement entropy difference $\Delta S(L)$ on transiting
  across the phase boundaries shown in Fig.~\ref{Fig:PD}. We use
  periodic boundaries with up to $L=64$ and work away from the
  multicritical point.  (a) The transition from MC to AC+MC yields
  $c\approx 3/2$ corresponding to an Ising transition coexisting with
  a gapless superfluid.  (b) The transition from MC to MI yields
  $c\approx 1$ and is consistent with a molecular KT transition.  (c)
  The transition from AC+MC to MI with $\epsilon_m=4$ yields $c\approx
  1$ and is consistent with an atomic KT transition. (d) The
  transition from AC+MC to MI with $\epsilon_m=-3.5$ appears to be
  compatible with the approach towards $c\approx 1$ with increasing
  $L$, although the finite-size effects are stronger than those in
  panel (c).  Panels (a) and (c) are adapted from
  Ref.~\cite{Ejima:ID}.}
\label{Fig:Delta}
\end{figure}

Having established good agreement between field theory and DMRG for
the MC and AC+MC phases, let us now examine the quantum phase
transition between them.  A key diagnostic in this 1D setting is the
central charge, $c$, which is a measure of the number of critical
degrees of freedom. This may be obtained from the entanglement
entropy. For a block of length $l$ in a periodic system of length $L$,
the von Neumann entropy is given by $S_L(l)=-{\rm Tr}_l(\rho_l\ln
\rho_l)$, where $\rho_l={\rm Tr}_{L-l}(\rho)$ is the reduced density
matrix. One obtains \cite{Holzey:Entropy,Calabrese:Entanglement}
\begin{equation}
S_L(l)=\frac{c}{3}\ln\left[\frac{L}{\pi}
\sin\left(\frac{\pi l}{L}\right)\right]+s_1+\ldots,
\label{ee}
\end{equation}
where $s_1$ is a non-universal constant and where the corrections are
small when the chord length is large
\cite{Calabrese:Parity,Cardy:Unusual,Calbrese:Universal,Xavier:Renyi,
Fagotti:Unusual,Eriksson:Corrections,Dalmonte:Estimating}.  As may be
seen in Fig.~\ref{Fig:Entang}(a), the numerically extracted central
charge of the MC phase yields $c=1$, as one would expect for a single
free boson, with coexisting gapped degrees of freedom;
the adjacent panel shows the same results plotted against the
conformal distance $\tilde l\equiv \ln[(L/\pi)\sin(\pi l/L)]$ in order
to yield a linear slope of $c/3$.  It may be seen from
Fig.~\ref{Fig:Entang}(c) that the AC+MC phase also has $c=1$. Note
that it is {\em not} $c=2$ as would be the case for two independent
Luttinger liquids. This reflects the coupled nature of the atomic and
molecular condensates in the AC+MC phase, with additional gapped Ising
degrees of freedom; the Feshbach term is relevant and drives the
${\mathbb Z}_2$ sector massive. Close to the MC to AC+MC transition,
where the anticipated Ising gap closes, one expects the central charge
to increase to $c=3/2$, due to {\em additional} critical Ising degrees
of freedom with $c=1/2$.  This is confirmed by our DMRG simulations in
Fig.~\ref{Fig:Entang}(b). Further support for this ${\mathbb Z}_2$
transition is obtained from the difference \cite{LK:Spreading},
\begin{equation}
\Delta S(L)\equiv S_L(L/2)-S_{L/2}(L/4)=\frac{c}{3}\ln(2)+\dots,
\end{equation}
as a function of $\epsilon_m$. For a given system size this displays a
peak, whose location coincides with the MC to AC+MC quantum phase
transition obtained via the vanishing of the single-particle gap,
$E_{1g}=0$, as shown in Fig.~\ref{Fig:PD}; see
Fig.~\ref{Fig:Delta}(a).
The evolution with increasing system size is consistent with the
passage towards $c=1$ in the superfluid phases, and $c=3/2$ in the
vicinity of the transition. This behavior may be contrasted with that
observed at the superfluid-MI transitions in Fig.~\ref{Fig:PD}, away
from the multicritical point. As may be seen in
Fig.~\ref{Fig:Delta}(b), in passing from the MC phase to the MI, the
central charge remains pinned at unity. This is consistent with a KT
transition for the molecules. Likewise, in passing from the AC+MC
phase to the MI, we find $c=1$ again; see
Fig.~\ref{Fig:Delta}(c). This is consistent with a KT transition for
the atoms. We have checked that this atomic KT behavior persists in
Fig.~\ref{Fig:PD} up to a value of $\epsilon_m=-3.5$; see
Fig.~\ref{Fig:Delta}(d). It is notable that the finite-size effects in
Fig.~\ref{Fig:Delta}(d) are much stronger than those in
Fig.~\ref{Fig:Delta}(c), although both are compatible with $c\approx
1$ at the MI transition. A detailed analysis of the multicritical
region in Fig.~\ref{Fig:PD} requires further investigation.

\section{Ising Scaling Regime}
\label{Sect:SR}
Having provided evidence for a ${\mathbb Z}_2$ quantum phase
transition occurring between the MC and AC+MC superfluids, we now
demonstrate how to extract both the Ising order parameter, $\langle
\phi\rangle$, and the Ising correlation length, $\xi$, in the presence
of the additional superfluid degrees of freedom with $c=1$.

\subsection{Ising Correlation Length}

\begin{figure}
\includegraphics[width=3.2in,clip=true]{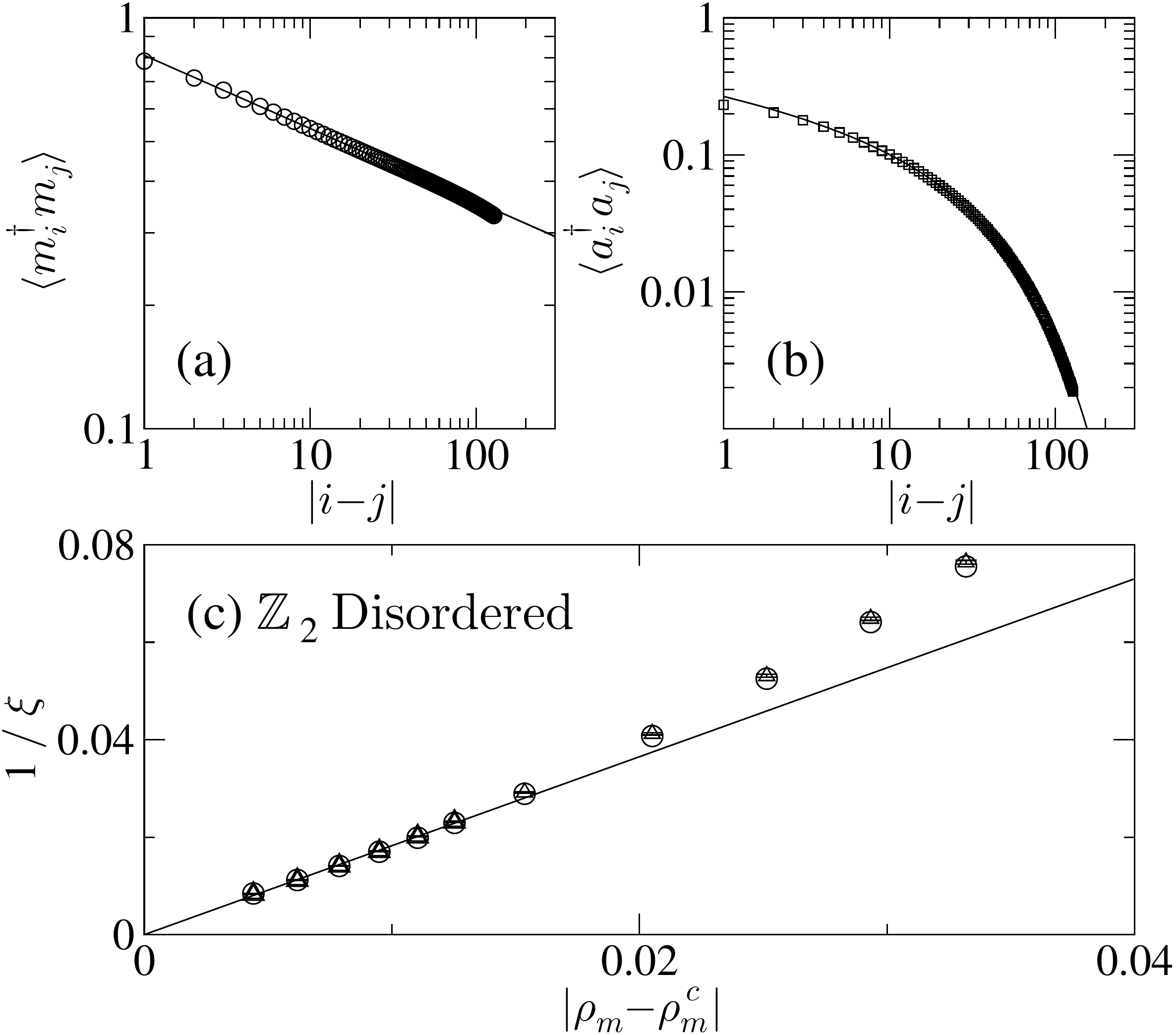}
\caption{DMRG results used to extract the Ising correlation length
  $\xi$ within the ${\mathbb Z}_2$ disordered MC phase with $L=256$
  and open boundaries. (a) With $\epsilon_m=-3.85$ the molecular
  correlation function $\langle m^\dagger_im_j\rangle\approx
  0.81|i-j|^{-0.18}$ decays as a power-law. (b) Using the previous
  exponent $\nu_m\approx 0.18$ we extract the Ising correlation length
  $\xi\approx 43.6$ from the exponential decay of $\langle
  a^\dagger_ia_j\rangle\sim |i-j|^{-\nu_m/4}{\rm K}_0(|i-j|/\xi)$. (c)
  Repeating the above procedure we plot $\xi^{-1}$ (circles) versus
  the departure of the molecular density $\rho_m$ from its value
  $\rho_m^c$ at the MC to AC+MC transition. Close to the transition
  the results are in good agreement with the Ising relation
  $\xi^{-1}\sim |\rho_m-\rho_m^c|^{\nu}$ with $\nu=1$. The triangles
  correspond to extracting $\xi$ directly from the ratio ${\mathcal
    R}(|i-j|)\equiv \langle a^\dagger_ia_j\rangle^4/\langle
  m^\dagger_im_j\rangle \sim [{\rm K}_0(|i-j|/\xi)]^4$.}
\label{Fig:Corrlen}
\end{figure}

The Ising correlation length, $\xi$, may be obtained from the atomic
and molecular correlation functions discussed in
Sec.~\ref{Sec:GF}. Within the ${\mathbb Z}_2$ disordered MC phase the
atomic correlations $\langle a^\dagger(x)a(0)\rangle\sim
x^{-\nu_m/4}{\rm K}_0\left(x/\xi\right)$ decay exponentially, whilst
the molecular correlations $\langle m^\dagger(x)m(0)\rangle\sim
x^{-\nu_m}$ decay as a power-law.  At a given point in parameter space
we may use the molecular Green's function to determine the exponent
$\nu_m$, and thereby extract the Ising correlation length from the
atomic correlations. This approach is outlined in
Fig.~\ref{Fig:Corrlen}. In the vicinity of an Ising quantum phase
transition one expects that $\xi^{-1}\sim |{\mathcal M}-{\mathcal
  M}_c|^\nu$ where $\nu=1$ is the Ising correlation length exponent
and ${\mathcal M}$ is a suitable mass scale parameterizing the
departure from criticality. Unfortunately, it is non-trivial to
express ${\mathcal M}$ in terms of the microscopic parameters of the
lattice model (\ref{atmolham}).  A naive analysis gives ${\mathcal
  M}\sim\kappa_0+\kappa_1 \rho_m+\kappa_2\sqrt{\rho_m}$, where
$\rho_m$ is the density of molecules, and $\kappa_0\sim \epsilon_a$,
$\kappa_1\sim U_{am}$, $\kappa_2\sim 2g$ are constants. Expanding the
square root according to
$\sqrt{\rho_m}\approx\sqrt{\rho_m^c}+(\rho_m-\rho_m^c)/\sqrt{\rho_m^c}$
suggests that sufficiently close to the Ising transition \be
\xi^{-1}\propto |\rho_m-\rho_m^c|.
\label{xirhom}
\ee 
This Ising behavior with $\nu=1$ 
is confirmed in Fig.~\ref{Fig:Corrlen}.

\subsection{Ising Order Parameter}
\begin{figure}
\includegraphics[width=3.2in,clip=true]{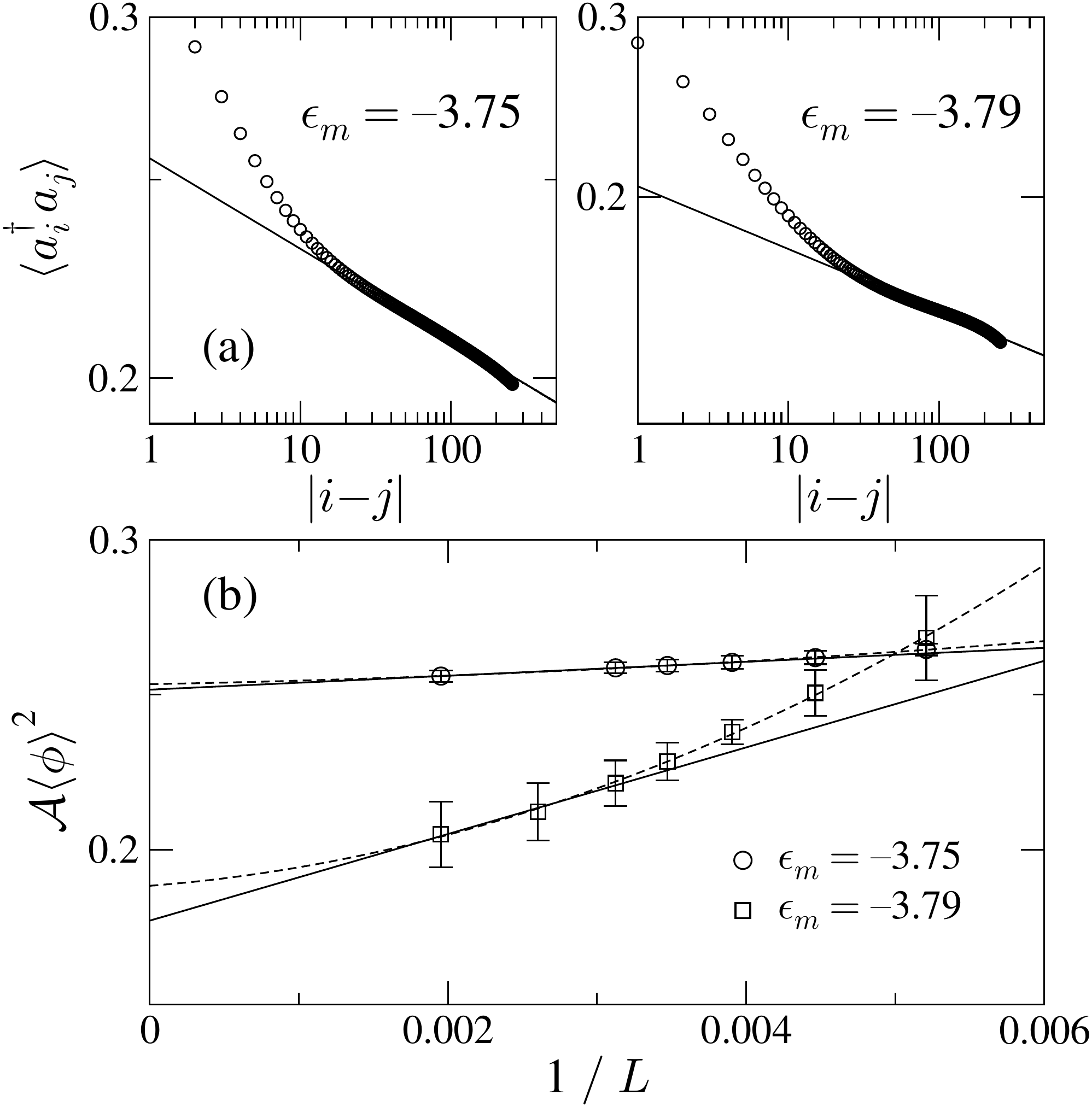}
\caption{(a) Atomic correlation functions within the AC+MC phase
  obtained by DMRG with up to $L=512$ and open boundaries. We set
  $U=0.7$ and consider $\epsilon_m=-3.75$ (left) and
  $\epsilon_m=-3.79$ (right).  A direct fit to Eq.~(\ref{aanop})
  yields ${\mathcal A}\langle\phi\rangle^2$ for each value of
  $\epsilon_m$, where $\langle \phi\rangle$ is the Ising order
  parameter and ${\mathcal A}$ is a non-universal constant prefactor.
  Changing the fitting interval gives an estimate of the error bars.
  (b) Extrapolation of ${\mathcal A}\langle \phi\rangle^2$ to the
  thermodynamic limit using linear extrapolation of the largest three
  system sizes is indicated by the solid line.  An estimate of the
  error bars in the thermodynamic limit is obtained by comparing to a
  quadratic fit of the data shown by the dashed line.  These results
  are plotted as a function of the molecular density in
  Fig.~\ref{Fig:phidens} in order to confirm Ising behavior with
  $\beta=1/8$.}
\label{Fig:lnaa}
\end{figure}

\begin{figure}
\includegraphics[width=3.2in,clip=true]{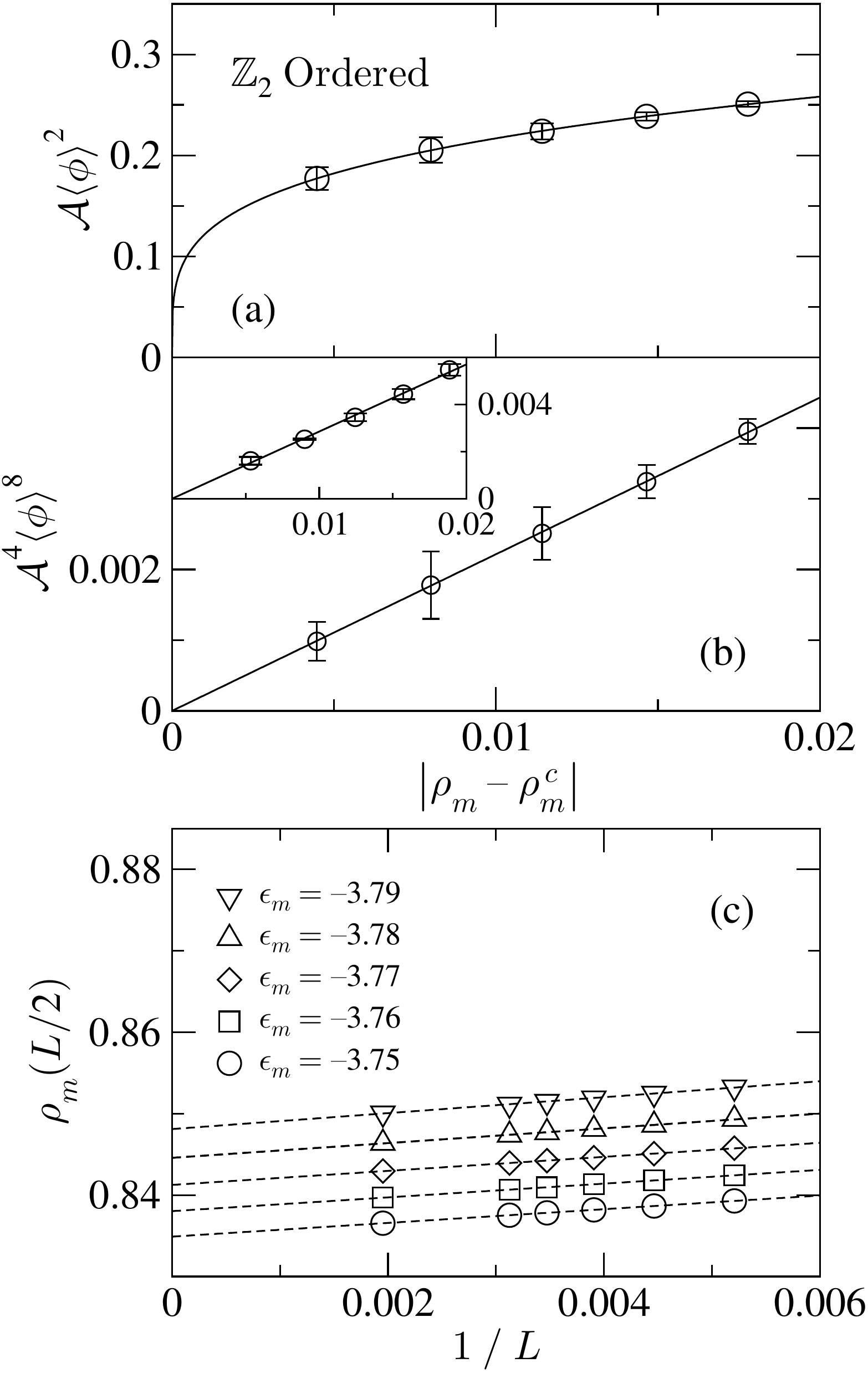}
\caption{DMRG results for the Ising order parameter in the ${\mathbb
    Z}_2$ ordered AC+MC phase with up to $L=512$ and $U=0.7$.  (a)
  Variation of the Ising order parameter squared ${\mathcal
    A}\langle\phi\rangle^2$ where ${\mathcal A}$ is a constant
  prefactor, versus the deviation of the molecular density $\rho_m$
  from its value $\rho_m^c$ at the MC to AC+MC quantum phase
  transition. (b) Variation of ${\mathcal A}^4\langle\phi\rangle^8$
  versus the molecular density difference. The results are in good
  agreement with the Ising magnetization relation $\langle
  \phi\rangle\sim |\rho_m-\rho_m^c|^{1/8}$ with $\beta=1/8$. The inset
  shows analogous results obtained from the plateau value of
  ${\mathcal R}(x)$ for $L=512$, as indicated in
  Fig.~\ref{Fig:Ratio}(c).  The error bars are estimated from the
  magnitude of $|{\mathcal R}(x=128, L=512)-{\mathcal R}(x=64,
  L=256)|$.  (c) Finite-size scaling of the thermodynamic molecular
  density used in panels (a) and (b).  Panels (a) and (b) are adapted
  from Ref.~\cite{Ejima:ID}.}
\label{Fig:phidens}
\end{figure}

In the ordered phase of the Ising model (\ref{lphi}) we have $\langle
\phi\rangle\sim |{\mathcal M}-{\mathcal M}_c|^{\beta}$ where
$\beta=1/8$ is the Ising magnetization critical exponent.  
From the discussion above one thus expects that
\be \langle
\phi\rangle\sim |\rho_m-\rho_m^c|^{1/8},
\label{phirhom}
\ee where $\rho_m$ is the density of molecules. In order to test the
validity of Eq.~(\ref{phirhom}), we must first extract the Ising order
parameter from a finite-size scaling analysis of the atomic
correlations. As follows from Eq.~(\ref{aaord}), within the ${\mathbb
  Z}_2$ ordered AC+MC phase one has
\begin{equation}
\langle a^\dagger(x)a(0)\rangle = {\mathcal A} \langle \phi\rangle^2x^{-\nu_m/4},
\label{aanop}
\end{equation} 
where ${\mathcal A}$ is a normalization amplitude.  In
Fig.~\ref{Fig:lnaa}(a) we show DMRG results for $\langle
a^\dagger(x)a(0)\rangle$ in the vicinity of the MC to AC+MC quantum
phase transition. A direct fit to Eq.~(\ref{aanop}) yields ${\mathcal
  A}\langle \phi\rangle^2$.
Repeating this procedure for different system sizes one obtains 
an estimate for ${\mathcal A}\langle\phi\rangle^2$ in the thermodynamic
limit; see Fig.~\ref{Fig:lnaa}(b). In Fig.~\ref{Fig:phidens} we show 
the variation of this order parameter with the molecular density. 
The results are in good agreement with the theoretical prediction
in Eq.~(\ref{phirhom}) and the Ising critical exponent $\beta=1/8$.

\subsection{Correlation Function Ratio}
In the above discussion we have extracted the Ising correlation length
and the Ising order parameter through a direct finite-size scaling
analysis of the atomic and molecular correlation functions. An
alternative approach is to consider the behavior of the ratio 
\begin{equation}
{\mathcal R}(x)\equiv \frac{\langle
    a^\dagger(x)a(0)\rangle^4}{\langle m^\dagger(x)m(0)\rangle},
\end{equation}
in analogy to the considerations of
Refs.~\cite{Lecheminant:Confinement,Capponi:Confinement} for the
confinement-deconfinement transition of Cooper pairs in 1D fermion
systems.

\begin{figure}[h]
\includegraphics[width=3.2in,clip]{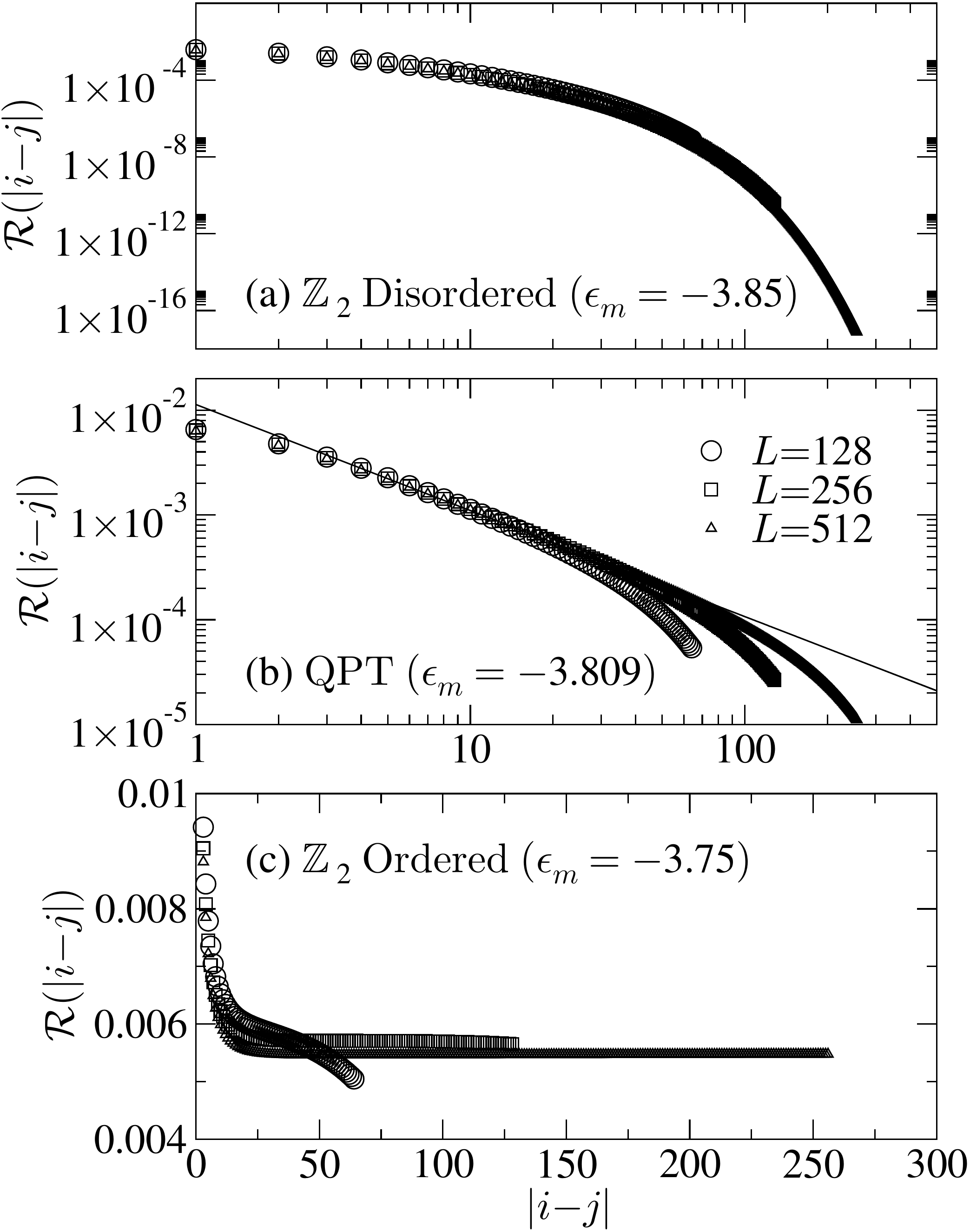}
\caption{Ratio ${\mathcal R}(|i-j|)\equiv \langle
  a^\dagger_ia_j\rangle^4/\langle m^\dagger_im_j\rangle$ of the atomic
  and molecular correlation functions with $U=0.7$.  (a) In the
  ${\mathbb Z}_2$ disordered MC phase with $\epsilon_m=-3.85$ the
  ratio ${\mathcal R}(|i-j|)\sim [{\rm K}_0(|i-j|/\xi)]^4$ exhibits
  exponential decay.  (b) In the vicinity of the MC to AC+MC quantum
  phase transition the ratio ${\mathcal R}(|i-j|)\sim 1/|i-j|$ decays
  with a universal power-law corresponding to the Ising critical
  exponent $\eta=1/4$. The line is a fit to ${\mathcal R}=A_0
  |i-j|^{A_1}$ over the interval $3\le |i-j|\le 48$ for $L=512$ with
  $A_0\approx 0.011$ and $A_1\approx -1.01$. (c) In the ${\mathbb
    Z}_2$ ordered AC+MC phase with $\epsilon_m=-3.75$ the ratio
  ${\mathcal R}(|i-j|)\sim \langle \phi\rangle^8$ exhibits a plateau
  corresponding to a non-zero Ising order parameter.}
\label{Fig:Ratio}
\end{figure}

In the ${\mathbb Z_2}$ disordered MC phase $\langle
a^\dagger(x)a(0)\rangle\sim x^{-\nu_m/4}{\rm K}_0\left(x/\xi\right)$
and $\langle m^\dagger(x)m(0)\rangle\sim x^{-\nu_m}$. It follows that
the power-law prefactors cancel out in this ratio:
\begin{equation}
{\mathcal R}(x)\sim \left[{\rm K}_0(x/\xi)\right]^4.
\label{rdis}
\end{equation}
As such, this ratio should exhibit exponential decay in the MC
phase. This is confirmed by our DMRG results in
Fig.~\ref{Fig:Ratio}(a).  
A direct fit to Eq.~(\ref{rdis}) yields the Ising correlation length shown by
the triangles in Fig.~\ref{Fig:Corrlen}.  

In the ${\mathbb Z}_2$ ordered AC+MC phase $\langle
m^\dagger(x)m(0)\rangle\sim x^{-\nu_m}$ and $\langle
a^\dagger(x)a(0)\rangle\sim \langle \phi(x)\phi(0)\rangle
x^{-\nu_m/4}$ where $\langle \phi(x)\phi(0)\rangle$ is given by
Eq.~(\ref{phicorr}). It follows that
\begin{equation}
{\mathcal R}(x)\sim \langle \phi\rangle^8\left[1+\pi^{-2}{\rm F}(x/\xi)\right]^4,
\end{equation}
where ${\rm F}(z)$ is given by Eq.~(\ref{Fdef}). At leading order
${\mathcal R}(x)\sim \langle \phi\rangle^8$ and one thus expects
${\mathcal R}(x)$ to develop a constant plateau that is proportional
to the Ising order parameter. This is confirmed by our DMRG results 
in Fig.~\ref{Fig:Ratio}(c).

In addition to these results for ${\mathcal R}(x)$ which are valid
within the superfluid phases, one may also explore the vicinity of the
quantum phase transition between them.  At the Ising critical point
$\langle\phi\rangle=0$ but \be \langle \phi(x)\phi(0)
\rangle\sim\left(\frac{a_0}{x}\right)^{\eta}, \label{phiphi} \ee
decays as a power-law where $\eta=1/4$ is the Ising pair correlation
exponent.  It follows from Eqs.~(\ref{amising}) and (\ref{phiphi})
that the atomic Green's function at criticality is given by \bea
\langle a^\dagger(x)a(0)\rangle&\sim& \langle \phi(x)\phi(0)\rangle\
\langle e^{-i\frac{\vartheta(x)}{2}}\
e^{i\frac{\vartheta(0)}{2}}\rangle\nn &\sim&
\left(\frac{a_0}{x}\right)^\frac{1}{4}
\left(\frac{a_0}{x}\right)^\frac{\nu_m}{4}.  \eea On passing from the
${\mathbb Z}_2$ ordered AC+MC phase towards the Ising quantum phase
transition, the power-law decay of the atomic Green's function is
therefore enhanced by $\eta=1/4$ due to the presence of additional
critical Ising degrees of freedom. It follows that
\begin{equation}
{\mathcal R}(x)\sim \frac{a_0}{x},
\end{equation}
exhibits universal power-law decay in the vicinity of the MC to AC+MC
quantum phase transition. This is confirmed by our DMRG results shown
in Fig.~\ref{Fig:Ratio}(b). This provides direct evidence for the
Ising correlation exponent, $\eta=1/4$.  These results demonstrate
that the ratio ${\mathcal R}(x)$ may be used to explore both the
critical and off-critical Ising behavior at the MC to AC+MC
transition. The characteristic signatures of ${\mathcal R}(x)$
parallel those observed in
Refs.~\cite{Lecheminant:Confinement,Capponi:Confinement} for the
confinement-deconfinement transition of Cooper pairs in 1D fermion
systems.
\section{Conclusions}
\label{Sect:Conc}

In this manuscript we have explored the phase diagram of bosons
interacting via Feshbach resonant pairing in a 1D optical lattice. We
have presented a wide variety of evidence in favor of an Ising quantum
phase transition separating distinct paired superfluids. We have also
provided a detailed characterization of these phases, including the
behavior close to the Mott insulating phase boundary. For the
investigated parameters, our DMRG results are consistent with an Ising
quantum phase transition approaching both a molecular KT transition
and an atomic KT transition. This is compatible with mean field theory
predictions for the continuum model in higher dimensions. However,
recent results for pairing phases in a 2D classical XY model suggest
the possibility that the Ising transition may over-extend beyond the
multicritical point \cite{Shi:Pairing}. In view of this possibility,
in a distinct but closely related model, it would be profitable to
explore the multicritical region in more detail. A clear verdict on
this issue for the present 1D quantum model requires further
analytical and numerical investigation and we will return to this
question in future work. It would also be interesting to explore the
phase diagram for a broader range of parameters, with a specific focus
on the choice of atomic species and experimental constraints. Even in
the presence of strong three body losses, the emergent phase diagram
may exhibit notable similarities
\cite{Daley:Three,Daley:Threeerratum,Diehl:Observability,Diehl:QFTI,Diehl:QFTII,Bonnes:Pair,Bonnes:Unbinding}.

\begin{acknowledgments}
  We are grateful to E. Altman, F. Assaad, S. Capponi, N. Cooper,
  S. Diehl, M. Garst, Z. Hadzibabic, A. James, E. Jeckelmann,
  J. Kj\"all, M. K\"ohl, A. Lamacraft, A. L{\"a}uchli, C. Lobo,
  J. Moore, N. Prokof'ev, A. Silver and M. Zaletel for helpful
  comments and discussions. MJB and BDS acknowledge EPSRC grant no.
  EP/E018130/1. FHLE by EP/I032487/1 and EP/D050952/1.  SE and HF
  acknowledge funding by the DFG through grant SFB 652. MH by DFG
  FG1162. Numerical calculations were performed at the URZ Greifswald.
\end{acknowledgments}

\appendix

\section{Hilbert Space Truncation}
\label{App:Trunc}
\begin{figure}
\includegraphics[width=3.2in,clip=true]{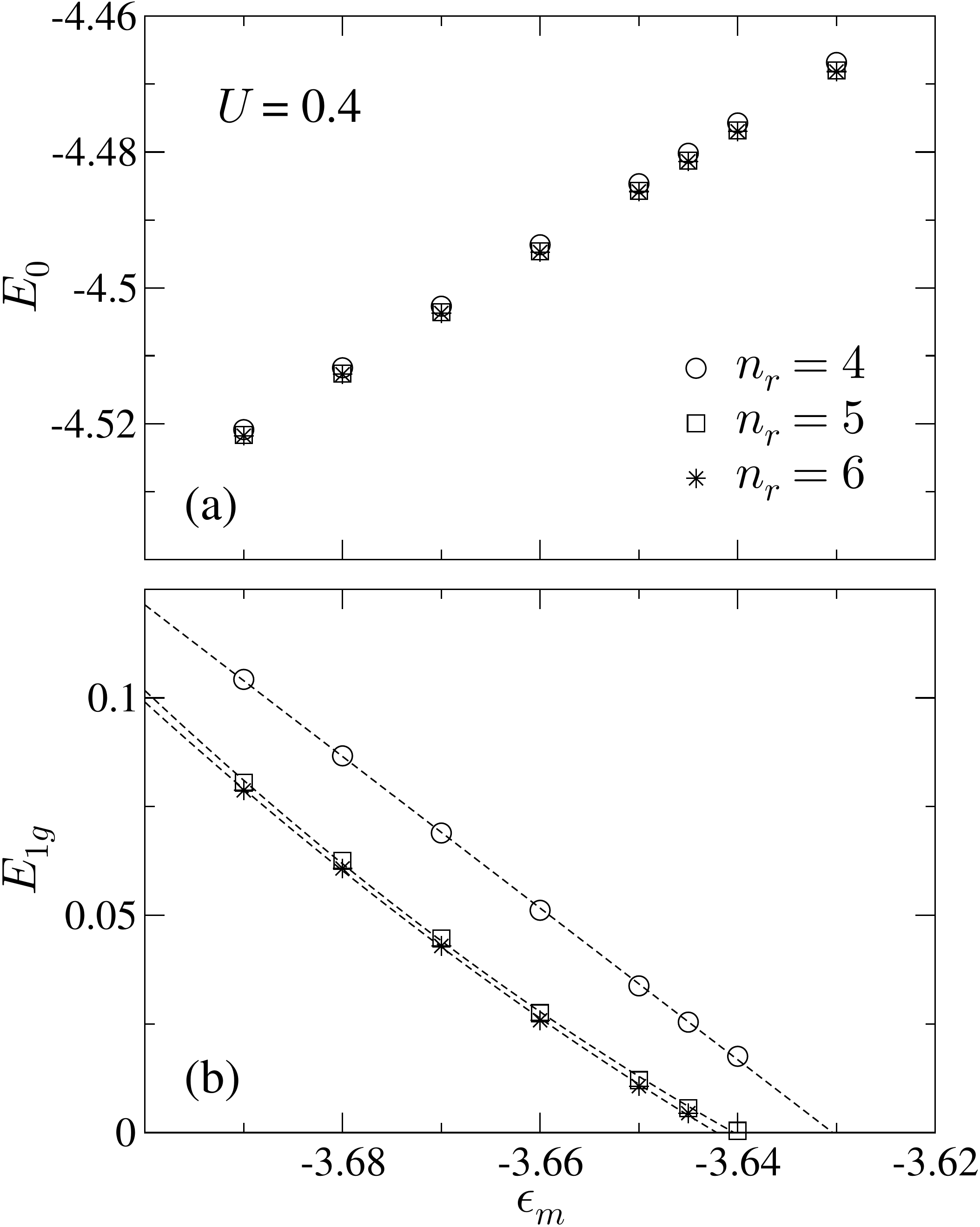}
\caption{DMRG results for a horizontal scan through Fig.~\ref{Fig:PD}
  with $1/U=2.5$. We consider up to $L=128$ and extrapolate to the
  thermodynamic limit. (a) Evolution of the ground state energy $E_0$
  with increasing local Hilbert space restriction, $n_r=n_a=n_m$. (b)
  Evolution of the excitation gap $E_{1g}$ with increasing $n_r$, showing
  very little change beyond $n_r=5$.}
\label{Fig:Entrunc}
\end{figure}

\begin{figure}
\includegraphics[width=3.2in,clip=true]{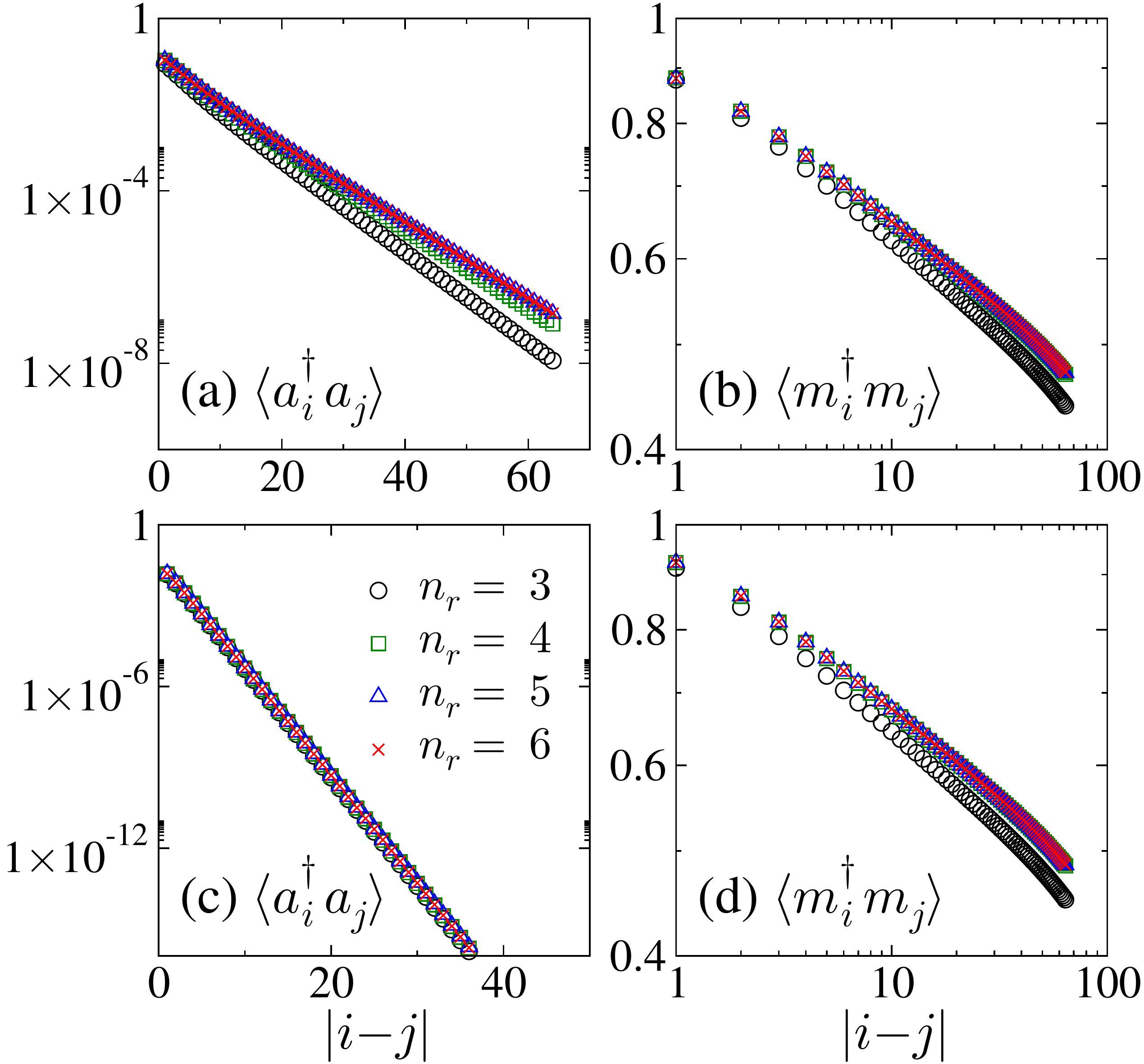}
\caption{(color online). DMRG results with $L=128$ and $U=0.5$ within
  the MC phase shown in Fig.~\ref{Fig:PD}. We show the evolution of
  the atomic and molecular correlation functions with increasing local
  Hilbert space restriction $n_r$. We set $\epsilon_m=-4$
  ($\epsilon_m=-6$) in the upper (lower) panels.}
\label{Fig:Truncmc}
\end{figure}

\begin{figure}
  \includegraphics[width=3.2in,clip=true]{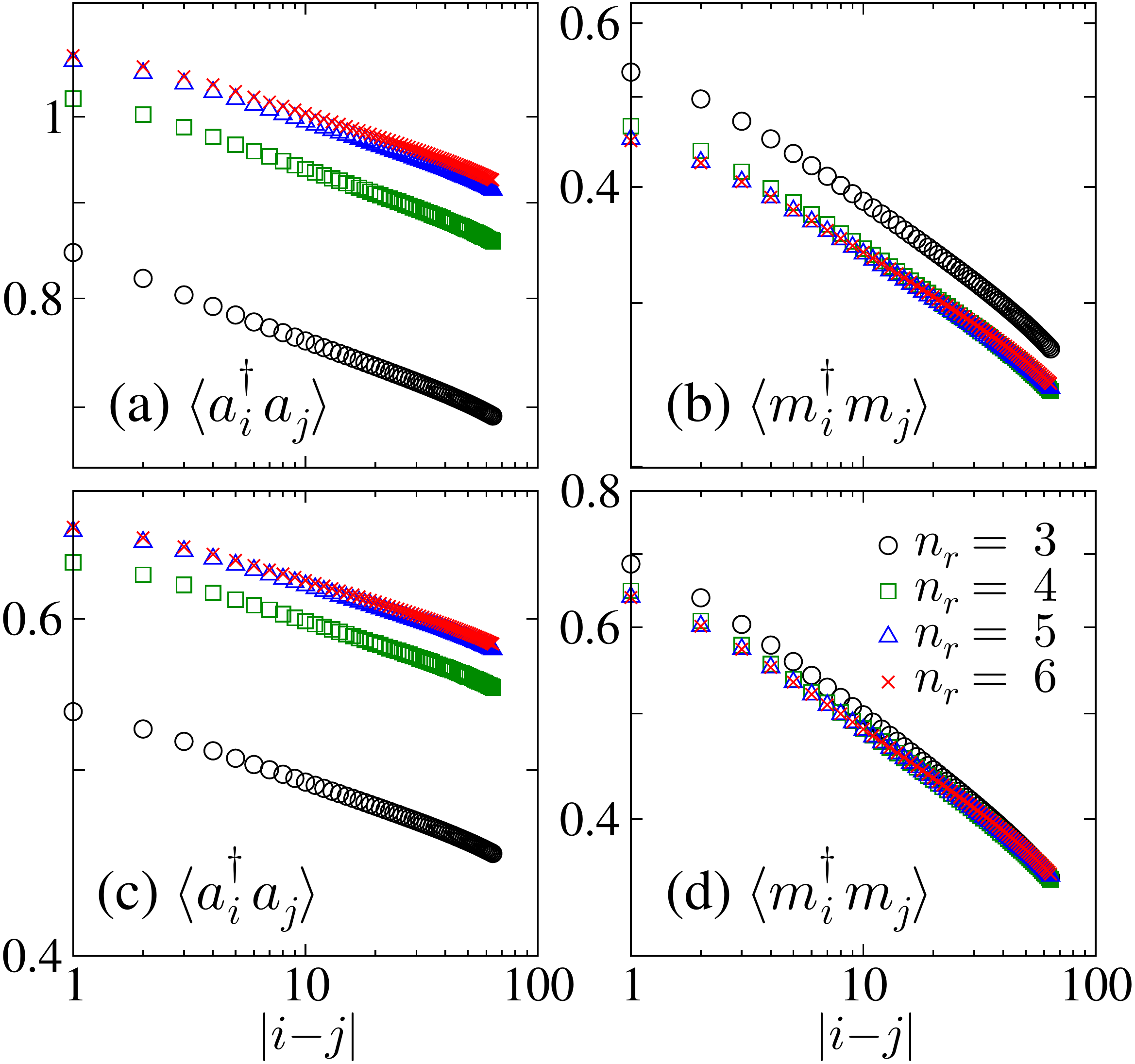}
  \caption{(color online). DMRG results with $L=128$ and $U=0.5$ within
    the AC+MC phase shown in Fig.~\ref{Fig:PD}. We show the evolution
    of the atomic and molecular correlation functions with increasing
    local Hilbert space restriction $n_r$. We set $\epsilon_m=-2$
    ($\epsilon_m=-3$) in the upper (lower) panels.}
\label{Fig:Truncacmc}
\end{figure}

Throughout the main text we truncate the local Hilbert space to allow
up to a maximum of $n_a=5$ atoms and $n_m=5$ molecules per site. In
the regime of large $t/U$, where inter-site hopping is strongly
favored, one should check the validity of this approximation. Here we
discuss the evolution of physical observables with increasing Hilbert
space restriction parameter, $n_r=n_a=n_m$. For the largest value of
$1/U=2.5$ used in Fig.~\ref{Fig:PD}, the results converge with
increasing $n_r$. For example, in Fig.~\ref{Fig:Entrunc}(a)
we show the evolution of the ground state energy $E_0$ with increasing
$n_r$.  The results show very little variation beyond $n_r=5$. Likewise,
in Fig.~\ref{Fig:Entrunc}(b) we monitor the excitation gap $E_{1g}$ with
increasing $n_r$. The data again show very little change beyond
$n_r=5$. The associated MC to AC+MC phase boundary in Fig.~\ref{Fig:PD}
is therefore robust to increasing $n_r$. In a similar fashion, in
Figs.~\ref{Fig:Truncmc} and \ref{Fig:Truncacmc},
we examine the evolution of the atomic and molecular correlation
functions.  The results show clear convergence in both the MC and
AC+MC phases. The excellent agreement between our DMRG results and
field theory predictions also lends {\em a postiori} support for this
level of Hilbert space restriction with $n_r=5$.

\end{document}